\documentclass[12pt]{article}

\author{Yu.~M.~Zinoviev
       \thanks{E-mail address: Yurii.Zinoviev@ihep.ru} \\[0.5cm]
        {\it Institute for High Energy Physics} \\
        {\it of National Research Center "Kurchatov Institute"} \\
        {\it Protvino, Moscow Region, 142280, Russia}}

\title{On higher spin cubic interactions in $d=3$}

\date{}

\begin{document}

\maketitle

\begin{abstract}
In this paper we elaborate on higher spin cubic interactions for
massless, massive and partially massless fields. We work in the gauge
invariant frame-like multispinor formalism, combining Lagrangian and
unfolded formulations.
\end{abstract}

\thispagestyle{empty}
\newpage
\setcounter{page}{1}

\tableofcontents

\section{Introduction}

Recently, a general classification of cubic interaction vertices for
massive higher spin fields in flat three dimensional space has been
developed in the light-cone formalism \cite{Met20} (see also
\cite{STT20}). As is well known, massless higher spin fields in $d=3$
do not have any local physical degrees of freedom and so they simply
do not exist in the light-cone formalism. At the same time,
interactions between massless and massive fields can lead to highly
non-trivial theories the best known example being the
Prokushlin-Vasiliev theory \cite{PV98,PSV99}. Besides, the possibility
to construct non-trivial interactions for partially massless fields
was discussed recently in \cite{GMS20}. In this paper we elaborate on
the higher spin cubic interactions for massless, massive and partially
massless fields. We work in the gauge invariant frame-like multispinor
formalism \cite{BSZ12a,BSZ12b,BSZ14a,Zin16} (see \cite{BSZ17a} for
review), combining Lagrangian and unfolded formulations.

One of the characteristic features of any gauge invariant description
for massive fields is the presence of the Stueckelberg fields. As it
has been shown in the metric-like formalism for $d \ge 4$
\cite{BDGT18} this fact leads to the following two important results.
At first, there always exist enough field redefinition which allows
one to transform any non-abelian interaction vertices (i.e. those that
deform both gauge transformations and the algebra) into the abelian
ones (i.e. those that deform gauge transformations, while the algebra
remains abelian). At second, using further (higher derivative) field
redefinition one can bring the vertex into the trivially gauge
invariant form (i.e. expressed completely in terms of the gauge
invariant objects of the free theory). Recently, on the two simple
concrete examples (massive spin-3/2 and massive spin-2) we considered
what happens if massless spin-2 field present in the system
\cite{KhZ21,KhZ21a}. In both cases we have found that there exist
abelian vertices which are not equivalent to any trivially gauge
invariant ones. Moreover, these vertices are necessary to reproduce
the minimal gravitational interactions. One of the aims of the current
work is to see what changes in $d=3$ where massless fields do not have
physical degrees of freedom and all their gauge invariant objects
vanish on-shell.

This paper is organized as follows. We begin in Section 2 with
massless higher spin fields and show how the frame-like multispinor
formalism allows easily reproduce the general classification of cubic
interaction vertices for the bosonic fields developed in
\cite{Mkr17,KM18} and also extend it including fermions. As a bonus,
in Section 3 we construct cubic interaction vertices for massless
higher spin supermultiplets corresponding to the simplest $(1,0)$
supersymmetry. These vertices correspond to the type II ones in the
Metsaev's classification of their four dimensional cousins
\cite{Met19a,Met19b} (see also \cite{KhZ20a,KhZ20b} and references
therein). Section 4 devoted to the massive fields. In the first two
subsections we construct a very simple gauge invariant Lagrangians for
bosons and fermions obtained from the initial ones by the (almost)
maximal gauge fixing. In both cases each Lagrangian can be associated
with the set of self-consistent unfolded equations. In subsection 4.3,
both in Lagrangian and unfolded approach, we investigate cubic
interactions for massive and massless fields. We consider all three
possibilities: two massive and one massless bosons; two massive
fermion and massless boson; massive boson and fermion and massless
fermion. In all three cases we found that the gauge invariance
implies the following relation on masses:
$$
M_1 - M_2 = (s_1-s_2)\lambda, \qquad \lambda = \sqrt{-\Lambda}
$$
which does not depend on the massless field spin. At last, in
subsection 4.4 we discuss cubic vertices for the three massive fields.
In Section 5 we consider partially massless fields. For the case with
maximal depth (the only one that has one physical degrees of freedom)
we consider their interaction with massless fields and show that both
Lagrangian and unfolded approach requires that spins of the partially
massless fields must be equal. For the partially massless fields with
non-maximal depth we show that our gauge invariant formalism can be
rewritten in the simple and elegant form suggested previously in
\cite{GMS20}.

\noindent
{\bf Notation and conventions} We mostly follow our review
\cite{BSZ17a}. All objects are forms having a number of completely
symmetric spinor indices which we denote $\alpha(n) =
(\alpha_1\alpha_2\dots\alpha_n)$. A background $AdS_3$ space is
described by the frame one-form $e^{\alpha(2)}$ and Lorentz covariant
derivative $D$ normalized so that
$$
D \wedge D \zeta^\alpha = - \lambda^2 E^\alpha{}_\beta \zeta^\beta
$$
Also we use two and three-forms defined as:
$$
e^{\alpha(2)} \wedge e^{\beta(2)} = \varepsilon^{\alpha\beta}
E^{\alpha\beta}, \qquad E^{\alpha(2)} \wedge e^{\beta(2)} =
\varepsilon^{\alpha\beta} \varepsilon^{\alpha\beta} E
$$
In the main text wedge product sign $\wedge$ will be omitted.

\section{Massless fields}

This section is devoted to the massless higher spin fields and their
interactions. This material is rather well known, but in our gauge
invariant formalism these fields serve as the building blocks for the
massive and partially massless ones.

\subsection{Free fields}

Massless spin-$s$ boson is described by a pair of one-forms
$\Omega^{\alpha(2s-2)}$ and $f^{\alpha(2s-2)}$ with $2s-2$ completely
symmetric local spinor indices which we collectively denote as 
$\alpha(2s-2)$. Free Lagrangian (three-form) for the massless field in
$AdS_3$ background looks like:
\begin{eqnarray}
(-1)^s {\cal L}_0 &=& (s-1) \Omega_{\alpha(2s-3)\beta} 
e^\beta{}_\gamma \Omega^{\alpha(2s-3)\gamma} + \Omega_{\alpha(2s-2)} D
f^{\alpha(2s-2)} \nonumber \\
 && + \frac{(s-1)}{4}\lambda^2 f_{\alpha(2s-3)\beta} e^\beta{}_\gamma
f^{\alpha(2s-3)\gamma}.
\end{eqnarray}
This Lagrangian is invariant under the following local gauge
transformations:
\begin{eqnarray}
\delta_0 \Omega^{\alpha(2s-2)} &=& D \eta^{\alpha(2s-2)} +
\frac{\lambda^2}{4} e^\alpha{}_\beta \xi^{\alpha(2s-3)\beta},
\nonumber \\
\delta_0 f^{\alpha(2s-2)} &=& D \xi^{\alpha(2s-2)} + e^\alpha{}_\beta
\eta^{\alpha(2s-3)\beta}, 
\end{eqnarray}
where parameters $\eta$, $\xi$ are zero-forms. One can construct two
gauge invariant two-forms (generalizing curvature and torsion in
gravity):
\begin{eqnarray}
{\cal R}^{\alpha(2s-2)} &=& D \Omega^{\alpha(2s-2)} + 
\frac{\lambda^2}{4} e^\alpha{}_\beta f^{\alpha(2s-3)\beta}, 
\nonumber \\
{\cal T}^{\alpha(2s-2)} &=& D f^{\alpha(2s-2)} + e^\alpha{}_\beta
\Omega^{\alpha(2s-3)\beta},
\end{eqnarray}
which we collectively call curvatures.

One of the specific properties of the three-dimensional case is the
possibility to separate variables. Indeed, let us consider the
following combinations of the fields and gauge parameters:
\begin{eqnarray}
\Omega_\pm^{\alpha(2s-2)} &=& \Omega^{\alpha(2s-2)} \pm 
\frac{\lambda}{2} f^{\alpha(2s-2)}, \nonumber \\
\eta_\pm^{\alpha(2s-2)} &=& \eta^{\alpha(2s-2)} \pm 
\frac{\lambda}{2} \xi^{\alpha(2s-2)}.
\end{eqnarray}
In the new variables we obtain:
\begin{equation}
(-1)^s {\cal L}_0 = \frac{1}{2\lambda} [ {\cal L}_+(\Omega_+) -
{\cal L}_-(\Omega_-) ],
\end{equation}
where
\begin{equation}
{\cal L}_\pm(\Omega_\pm) = \Omega_{\pm,\alpha(2s-2)} D
\Omega_\pm^{\alpha(2s-2)} \pm (s-1)\lambda 
\Omega_{\pm,\alpha(2s-3)\beta} e^\beta{}_\gamma
\Omega_\pm^{\alpha(2s-3)\gamma}
\end{equation}
Moreover, each Lagrangian ${\cal L}_\pm$ is invariant under its own
gauge transformations:
\begin{equation}
\delta \Omega_\pm^{\alpha(2s-2)} = D \eta_\pm^{\alpha(2s-2)}
 \pm \frac{\lambda}{2} e^\alpha{}_\beta \eta_\pm^{\alpha(2s-3)\beta}.
\end{equation}
The same procedure works for the gauge invariant curvatures as well:
\begin{equation}
{\cal R}_\pm^{\alpha(2s-2)} = D \Omega_\pm^{\alpha(2s-2)}
 \pm \frac{\lambda}{2} e^\alpha{}_\beta 
\Omega_\pm^{\alpha(2s-3)\beta}.
\end{equation}
In what follows we work with the $+$  components only omitting the $+$
sign.

Massless spin-$(s+1/2)$ fermion is described by one-form 
$\Phi^{\alpha(2s-1)}$. Free Lagrangian in $AdS_3$ looks like:
\begin{equation}
\frac{(-1)^s}{i} {\cal L}_0 = \frac{1}{2} \Phi_{\alpha(2s-1)} D
\Phi^{\alpha(2s-1)} \pm \frac{(2s-1)}{4}\lambda 
\Phi_{\alpha(2s-2)\beta} e^\beta{}_\gamma \Phi^{\alpha(2s-2)\gamma},
\end{equation}
which is invariant under the following local gauge transformations:
\begin{equation}
\delta_0 \Phi^{\alpha(2s-1)} = D \zeta^{\alpha(2s-1)} \pm
\frac{\lambda}{2} e^\alpha{}_\beta \zeta^{\alpha(2s-2)\beta}.
\end{equation}
Here one can also construct a gauge invariant curvature:
\begin{equation}
{\cal F}^{\alpha(2s-1)} = D \Phi^{\alpha(2s-1)} \pm
\frac{\lambda}{2} e^\alpha{}_\beta \Phi^{\alpha(2s-2)\beta}
\end{equation}
For the fermions we also use the $+$ sign case.

\subsection{Cubic vertices}

In the metric-like formalism a complete classification for the bosonic
cubic vertices were elaborated in \cite{Mkr17,KM18}. In this
subsection we show how these results appear in the frame-like
formalism and extend them including fermionic fields.

Let us begin with the vertex for three bosons with spins $s_{1,2,3}$.
The most general ansatz has the form:
\begin{equation}
{\cal L}_1 = g \Omega_1^{\alpha(\hat{s}_3)\beta(\hat{s}_2)} 
\Omega_2^{\gamma(\hat{s}_1)}{}_{\alpha(\hat{s}_3)} 
\Omega_{3,\beta(\hat{s}_2)\gamma(\hat{s}_1)}.
\end{equation}
Here the $\hat{s}_i$ must satisfy
\begin{equation}
\hat{s}_2+\hat{s}_3 = 2s_1-2, \qquad 
\hat{s}_1+\hat{s}_3 = 2s_2-2, \qquad 
\hat{s}_1+\hat{s}_2 = 2s_3-2
\end{equation}
and this gives
\begin{equation}
\hat{s}_3 = s_1+s_2-s_3-1, \qquad
\hat{s}_2 = s_1+s_3-s_2-1 , \qquad
\hat{s}_1 = s_2+s_3-s_1-1.
\end{equation}
From the requirement $\hat{s}_i \ge 0$ immediately follows that the
three spins must satisfy a so-called strict triangular inequality
$s_i < s_{i+1} + s_{i+2}$. Note also that all $\hat{s}_i$ are
simultaneously odd or even, so that
\begin{equation}
(-1)^{\hat{s}_1} = (-1)^{\hat{s}_2} = (-1)^{\hat{s}_3}.
\end{equation}
Consider $\eta_3$-transformations as an example:
\begin{eqnarray}
\delta_0 {\cal L}_1 &=& g \Omega_1^{\alpha(\hat{s}_3)\beta(\hat{s}_2)}
\Omega_2^{\gamma(\hat{s}_1)}{}_{\alpha(\hat{s}_3)} 
[ D \eta_{\beta(\hat{s}_2)\gamma(\hat{s}_1)} - \frac{\lambda}{2}
e_\beta{}^\delta \eta_{\beta(\hat{s}_2-1)\delta\gamma(\hat{s}_1)}
- \frac{\lambda}{2} e_\gamma{}^\delta
\eta_{\beta(\hat{s}_2)\gamma(\hat{s}_1-1)\delta} ] \nonumber \\
 &=& - g D \Omega_1^{\alpha(\hat{s}_3)\beta(\hat{s}_2)}
\Omega_2^{\gamma(\hat{s}_1)}{}_{\alpha(\hat{s}_3)}
\eta_{\beta(\hat{s}_2)\gamma(\hat{s}_1)} 
 + g \Omega_1^{\alpha(\hat{s}_3)\beta(\hat{s}_2)}
D \Omega_2^{\gamma(\hat{s}_1)}{}_{\alpha(\hat{s}_3)}
\eta_{\beta(\hat{s}_2)\gamma(\hat{s}_1)} \nonumber \\
 && - \frac{g\lambda}{2} \hat{s}_2 
\Omega_1^{\alpha(\hat{s}_3)\beta(\hat{s}_2-1)\rho}
\Omega_2^{\gamma(\hat{s}_1)}{}_{\alpha(\hat{s}_3)}
e_\rho{}^\delta \eta_{\beta(\hat{s}_2-1)\delta\gamma(\hat{s}_1)}
\nonumber \\
 && - \frac{g\lambda}{2} \hat{s}_1
\Omega_1^{\alpha(\hat{s}_3)\beta(\hat{s}_2)}
\Omega_2^{\gamma(\hat{s}_1-1)\rho}{}_{\alpha(\hat{s}_3)}
e_\rho{}^\delta \eta_{\beta(\hat{s}_2)\gamma(\hat{s}_1-1)\delta}.
\end{eqnarray}
To compensate for these variations we introduce the following 
corrections to the gauge transformations:
\begin{eqnarray}
\delta_1 \Omega_1^{\alpha(2s_1-2)} &=& g_1 
\Omega_2^{\alpha(\hat{s}_3)\gamma(\hat{s}_1)}
\eta_3^{\alpha(\hat{s}_2)}{}_{\gamma(\hat{s}_1)} \nonumber \\
\delta_1 \Omega_2^{\alpha(2s_2-2)} &=& g_2
\Omega_1^{\alpha(\hat{s}_3)\beta(\hat{s}_2)}
\eta_3^{\alpha(\hat{s}_1)}{}_{\beta(\hat{s}_2)}.
\end{eqnarray}
They produce:
\begin{eqnarray}
\delta_1 {\cal L}_0 &=& 2g_1 \frac{(2s_1-2)!}{\hat{s}_2!\hat{s}_3!} 
D \Omega^{\alpha(\hat{s}_3)\beta(\hat{s}_2)}
\Omega_2^{\gamma(\hat{s}_1)}{}_{\alpha(\hat{s}_3)}
\eta_{\beta(\hat{s}_2)\gamma(\hat{s}_1)} \nonumber \\
 && + 2g_2 (-1)^{\hat{s}_1} \frac{(2s_2-2)!}{\hat{s}_1!\hat{s}_3!} 
\Omega_1^{\alpha(\hat{s}_3)\beta(\hat{s}_2)}
D \Omega_2^{\gamma(\hat{s}_1)}{}_{\alpha(\hat{s}_3)}
\eta_{\gamma(\hat{s}_1)\beta(\hat{s}_2)} \nonumber \\
 && + g_1\lambda \frac{(2s_1-2)!}{\hat{s}_3!(\hat{s}_2-1)!}
\Omega_{1,\alpha(\hat{s}_3)\beta(\hat{s}_2-1)\delta} e^\delta{}_\rho
\Omega^{\alpha(\hat{s}_3)\gamma(\hat{s}_1)}
\eta^{\beta(\hat{s}_2-1)\rho}{}_{\gamma(\hat{s}_1)} \nonumber \\
 && + g_2\lambda \frac{(2s_2-2)!}{\hat{s}_3!(\hat{s}_1-1)!}
\Omega_{2,\alpha(\hat{s}_3)\gamma(\hat{s}_1-1)\delta} e^\delta{}_\rho
\Omega_1^{\alpha(\hat{s}_3)\beta(\hat{s}_2)}
\eta^{\gamma(\hat{s}_1-1)\rho}{}_{\beta(\hat{s}_2)} \nonumber \\
 && + g_1\lambda \frac{(2s_1-2)!}{(\hat{s}_3-1)!\hat{s}_2!}
\Omega_{1,\alpha(\hat{s}_3-1)\beta(\hat{s}_2)\delta} e^\delta{}_\rho
\Omega_2^{\alpha(\hat{s}_3-1)\rho\gamma(\hat{s}_1)}
\eta^{\beta(\hat{s}_2)}{}_{\gamma(\hat{s}_1)} \nonumber \\
 && + g_2\lambda \frac{(2s_2-2)!}{(\hat{s}_3-1)!\hat{s}_1!} 
\Omega_{2,\alpha(\hat{s}_3-1)\gamma(\hat{s}_1)\delta} e^\delta{}_\rho
\Omega_1^{\alpha(\hat{s}_3-1)\rho\beta(\hat{s}_2)}
\eta^{\gamma(\hat{s}_1)}{}_{\beta(\hat{s}_2)}
\end{eqnarray}
This gives:
\begin{equation}
g_1 = - \frac{(\hat{s}_2)!(\hat{s}_3)!}{2(\hat{s}_2+\hat{s}_3)!}g,
\qquad g_2 = (-1)^{\hat{s}_1} 
\frac{(\hat{s}_1)!(\hat{s}_2)!}{2(\hat{s}_1+\hat{s}_2)!} g
\end{equation}
Similarly for two other gauge transformations. All corrections
correspond to the following deformations of the gauge invariant
curvatures:
\begin{eqnarray}
\Delta {\cal R}_1^{\alpha(2s_1-2)} &=& g_1 
\Omega_2^{\alpha(\hat{s}_3)\gamma(\hat{s}_1)}
\Omega_3^{\alpha(\hat{s}_2)}{}_{\gamma(\hat{s}_1)} \nonumber \\
\Delta {\cal R}_2^{\alpha(2s_2-2)} &=& g_2 
\Omega_1^{\alpha(\hat{s}_3)\beta(\hat{s}_2)}
\Omega_3^{\alpha(\hat{s}_1)}{}_{\beta(\hat{s}_2)} \\
\Delta {\cal R}_3^{\alpha(2s_3-2)} &=& g_3
\Omega_1^{\alpha(\hat{s}_2)\beta(\hat{s}_3)}
\Omega_2^{\alpha(\hat{s}_1)}{}_{\beta(\hat{s}_3)} \nonumber
\end{eqnarray}
The vertex constructed does not have any definite parity. So to
compare our results with the ones in the metric-like formalism,
let us temporally restore components with $-$ sign and consider
\begin{equation}
{\cal L}_1 = g_+ \Omega_{+1}^{\alpha(\hat{s}_3)\beta(\hat{s}_2)} 
\Omega_{+2}^{\gamma(\hat{s}_1)}{}_{\alpha(\hat{s}_3)} 
\Omega_{+3,\beta(\hat{s}_2)\gamma(\hat{s}_1)}
 + g_- \Omega_{-1}^{\alpha(\hat{s}_3)\beta(\hat{s}_2)} 
\Omega_{-2}^{\gamma(\hat{s}_1)}{}_{\alpha(\hat{s}_3)} 
\Omega_{-3,\beta(\hat{s}_2)\gamma(\hat{s}_1)}
\end{equation}
We obtain:
\begin{eqnarray}
{\cal L}_1 
 &=& (g_+ + g_-) [ \Omega_1^{\alpha(\hat{s}_3)\beta(\hat{s}_2)} 
\Omega_2^{\gamma(\hat{s}_1)}{}_{\alpha(\hat{s}_3)} 
\Omega_{3,\beta(\hat{s}_2)\gamma(\hat{s}_1)} \nonumber \\
 && \qquad \qquad + \frac{\lambda^2}{4} 
( \Omega_1^{\alpha(\hat{s}_3)\beta(\hat{s}_2)} 
f_2^{\gamma(\hat{s}_1)}{}_{\alpha(\hat{s}_3)} 
f_{3,\beta(\hat{s}_2)\gamma(\hat{s}_1)}
+ f_1^{\alpha(\hat{s}_3)\beta(\hat{s}_2)} 
f_2^{\gamma(\hat{s}_1)}{}_{\alpha(\hat{s}_3)} 
\Omega_{3,\beta(\hat{s}_2)\gamma(\hat{s}_1)} \nonumber \\
 && \qquad \qquad 
 + f_1^{\alpha(\hat{s}_3)\beta(\hat{s}_2)} 
\Omega_2^{\gamma(\hat{s}_1)}{}_{\alpha(\hat{s}_3)} 
f_{3,\beta(\hat{s}_2)\gamma(\hat{s}_1)}) ] \nonumber \\
 && + (g_+-g_-)\frac{\lambda}{2} [ 
( \Omega_1^{\alpha(\hat{s}_3)\beta(\hat{s}_2)} 
\Omega_2^{\gamma(\hat{s}_1)}{}_{\alpha(\hat{s}_3)} 
f_{3,\beta(\hat{s}_2)\gamma(\hat{s}_1)} 
 + \Omega_1^{\alpha(\hat{s}_3)\beta(\hat{s}_2)} 
f_2^{\gamma(\hat{s}_1)}{}_{\alpha(\hat{s}_3)} 
\Omega_{3,\beta(\hat{s}_2)\gamma(\hat{s}_1)} \nonumber \\
 && \qquad + f_1^{\alpha(\hat{s}_3)\beta(\hat{s}_2)} 
\Omega_2^{\gamma(\hat{s}_1)}{}_{\alpha(\hat{s}_3)} 
\Omega_{3,\beta(\hat{s}_2)\gamma(\hat{s}_1)}) 
  + \frac{\lambda^2}{4} f_1^{\alpha(\hat{s}_3)\beta(\hat{s}_2)} 
f_2^{\gamma(\hat{s}_1)}{}_{\alpha(\hat{s}_3)} 
f_{3,\beta(\hat{s}_2)\gamma(\hat{s}_1)} ]
\end{eqnarray}
Thus we obtain two independent vertices. The first one has three
derivatives (and one derivative tail); it is parity even/odd when sum
of the spins is odd/even. The second one has two derivatives (and zero
derivative tail); it is parity even/odd when sum of the spins is 
even/odd. These results are in complete agreement with the
classification given in \cite{Mkr17,KM18}.

In the multispinor formalism it is easy to extend these results to
include fermions. Let us consider the case for one boson with spin
$s_1$ and two fermions with spins $s_2+1/2,s_3+1/2$. Then the vertex
appears to be:
\begin{equation}
{\cal L}_1 = g \Omega_1^{\alpha(\hat{s}_3)\beta(\hat{s}_2)} 
\Phi_2^{\gamma(\hat{s}_1+1)}{}_{\alpha(\hat{s}_3)} 
\Phi_{3,\beta(\hat{s}_2)\gamma(\hat{s}_1+1)}
\end{equation}
where $\hat{s}_i$ are the same as before. 

In the frame-like formalism it is impossible to construct any vertices
higher than cubic ones. Thus to construct consistent model one has to
find such collection of massless fields and their cubic vertices that
the gauge (super)algebra closes. One more specific properties of three
dimensional case is that there exist models with finite number of
fields. The most simple and rather popular with the algebra $SL(n)$
\cite{CFPT10} describes all integer spins $2,3, \dots, n$. It is
possible to truncate these models to even spins $2,4,\dots,2n$ only
\cite{CLW12}, the algebra being $Sp(2n)$. May be the most simple
examples including fermions \cite{Zin14a} correspond to superalgebras
$OSp(1,2n)$ and describe even spins $2,4,\dots,2n$ and one
half-integer
spin $n+1/2$. The case $n=1$ is just $(1,0)$ supergravity, while $n=2$
describes $AdS_3$ hypergravity with spin-2 $\omega^{\alpha(2)}$,  
spin-4 $\Sigma^{\alpha(6)}$ and spin-5/2 $\Psi^{\alpha(3)}$. Bosonic
part of the vertex
\begin{equation}
 {\cal L}_{1b} = \frac{g}{3} \omega_\alpha{}^\beta
\omega_\beta{}^\gamma \omega_\gamma{}^\alpha + 3g
\Sigma_{\alpha(5)\beta} \omega^\beta{}_\gamma \Sigma^{\alpha(5)\gamma}
+ \frac{10\tilde{g}}{3} \Sigma_{\alpha(3)\beta(3)}
\Sigma^{\beta(3)}{}_{\gamma(3)} \Sigma^{\alpha(3)\gamma(3)}
\end{equation}
where $\tilde{g} = \sqrt{10}g$, contains $(2-2-2)$, $(2-4-4)$ and
$(4-4-4)$ subvertices, while fermionic part
\begin{equation}
{\cal L}_{1f} = \frac{3ig}{2} \Psi_{\alpha(2)\beta}
\omega^\beta{}_\gamma \Psi^{\alpha(2)\gamma} + 
\frac{3i\tilde{g}}{2} \Psi_{\alpha(3)} 
\Sigma^{\alpha(3)}{}_{\beta(3)} \Psi^{\beta(3)}
\end{equation}
contains $(2-5/2-5/2)$ and $(4-5/2-5/2)$ ones. All of them follow the
general pattern described above.

\section{Massless supermultiplets}

In four dimensions the complete classification of cubic vertices for
the massless higher spin supermultiplets were developed in the
light-cone formalism by Metsaev \cite{Met19a,Met19b} (see also
\cite{KhZ20a,KhZ20b} and references therein). In this section we show
that in three dimensions we can construct cubic vertices corresponding
to type II ones in Metsaev's classification.

\subsection{Free supermultiplets}

We work with the supermultiplets for the simplest $(1,0)$ global
superalgebra. Recall that in $AdS_3$ by global supertransformations we
mean such that their spinor parameter satisfies
\begin{equation}
D \zeta^\alpha + \frac{\lambda}{2} e^\alpha{}_\beta \zeta^\beta = 0.
\end{equation}
There exist two massless supermultiplets: with integer superspin 
$(s,s+1/2)$ and with half-integer one $(s,s-1/2)$. All we need is
their explicit supertransformations such that the sum of the free
bosonic and fermionic Lagrangians is invariant and the superalgebra
closes on-shell (for more details see \cite{BSZ17a} and references
therein).

\noindent
{\bf Integer superspin} $(s,s+1/2)$
\begin{eqnarray}
\delta \Omega^{\alpha(2s-2)} &=& i\sqrt{2s-1}\lambda
\Phi^{\alpha(2s-2)\beta} \zeta_\beta \nonumber \\
\delta \Phi^{\alpha(2s-1)} &=& \frac{1}{\sqrt{2s-1}}  
\Omega^{\alpha(2s-2)} \zeta^\alpha
\end{eqnarray}
{\bf Half-integer superspin} $(s,s-1/2)$
\begin{eqnarray}
\delta \Omega^{\alpha(2s-2)} &=& \frac{i\lambda}{\sqrt{2s-2}}  
\Psi^{\alpha(2s-3)} \zeta^\alpha \nonumber \\
\delta \Psi^{\alpha(2s-3)} &=& \sqrt{2s-2} \Omega^{\alpha(2s-3)\beta}
\zeta_\beta
\end{eqnarray}

\subsection{Cubic vertices}

General procedure we use here is the same as we have already used in
four dimensions \cite{KhZ20b}. Namely, having in our disposal three
supermultiplets, i.e. three bosonic and three fermionic fields, we can
construct four elementary vertices: one purely bosonic and three with
fermions (schematically)
$$
{\cal L}_1 \sim g_0 \Omega_1\Omega_2\Omega_3 +
g_1 \Omega_1\Phi_2\Phi_3 + g_2 \Phi_1\Omega_2\Phi_3
+ g_3 \Phi_1\Phi_2\Omega_3.
$$
Thus we just have to adjust the four coupling constants so that the
vertex be invariant under the global supertransformations. There are
three type II vertices in Metsaev's classification: type IIa with
three half-integer superspins and type IIb,c with two integer and one
half-integer superspins. The difference for type IIb and type IIc
comes from the fact that the number of derivatives in the four
dimensional cubic vertices strongly depends on which fields has lowest
spin. In three dimensions the spin ordering does not matter and this
leaves us with just two possibilities, which we consider in turn.

\noindent
{\bf Two integer and one half-integer superspins}
$(s_1,s_1+1/2)$, $(s_2,s_2+1/2)$, $(s_3,s_3-1/2)$ \\
In this case a candidate for the supersymmetric vertex looks like:
\begin{eqnarray}
{\cal L}_1 &=& g_0 \Omega_1^{\alpha(\hat{s}_3)\beta(\hat{s}_2)}
\Omega_2^{\gamma(\hat{s}_1)}{}_{\alpha(\hat{s}_3)}
\Omega_{3,\beta(\hat{s}_2)\gamma(\hat{s}_1)} \nonumber \\
 && + ig_1\lambda \Omega_1^{\alpha(\hat{s}_3+1)\beta(\hat{s}_2-1)}
\Phi_2^{\gamma(\hat{s}_1)}{}_{\alpha(\hat{s}_3+1)}
\Psi_{3,\beta(\hat{s}_2-1)\gamma(\hat{s}_1)} \nonumber \\
 && + ig_2\lambda \Phi_1^{\alpha(\hat{s}_3+1)\beta(\hat{s}_2)}
\Omega_2^{\gamma(\hat{s}_1-1)}{}_{\alpha(\hat{s}_3+1)}
\Psi_{3,\beta(\hat{s}_2)\gamma(\hat{s}_1-1)} \nonumber \\
 && + ig_3\lambda \Phi_1^{\alpha(\hat{s}_3+1)\beta(\hat{s}_2)}
\Phi_2^{\gamma(\hat{s}_1)}{}_{\alpha(\hat{s}_3+1)}
\Omega_{3,\beta(\hat{s}_2)\gamma(\hat{s}_1)}
\end{eqnarray}
Note that in this case we must have $\hat{s}_{1,2} \ge 1$, 
$\hat{s}_3 \ge 0$. Let us calculate variations of the vertex
containing two bosons and one fermion:
\begin{eqnarray*}
\frac{1}{i\lambda}\Delta_1 &=& [ \frac{g_0\hat{s}_2}{\sqrt{2s_3-2}} 
- \frac{g_1(\hat{s}_3+1)}{\sqrt{2s_2-1}}  ]
\Omega_1^{\alpha(\hat{s}_3)\beta(\hat{s}_2-1)\delta}
\Omega_2^{\gamma(\hat{s}_1)}{}_{\alpha(\hat{s}_3)}
\Psi_{3,\beta(\hat{s}_2-1)\gamma(\hat{s}_1)} \zeta_\delta \\
 && + [ \frac{g_0\hat{s}_1}{\sqrt{2s_3-2}} 
 + \frac{g_2(\hat{s}_3+1)}{\sqrt{2s_1-1}} ]
\Omega_1^{\alpha(\hat{s}_3)\beta(\hat{s}_2)}
\Omega_2^{\gamma(\hat{s}_1-1)\delta}{}_{\alpha(\hat{s}_3)}
\Psi_{3,\beta(\hat{s}_2)\gamma(\hat{s}_1-1)} \zeta_\delta \\
 && - [ \frac{g_1\hat{s}_1}{\sqrt{2s_2-1}}
 + \frac{g_2\hat{s}_2}{\sqrt{2s_1-1}} ]
\Omega_1^{\alpha(\hat{s}_3+1)\beta(\hat{s}_2-1)}
\Omega_2^{\gamma(\hat{s}_1-1)}{}_{\alpha(\hat{s}_3+1)}
\Psi_{3,\beta(\hat{s}_2-1)\gamma(\hat{s}_1-1)\delta} \zeta^\delta \\
 && + [ g_0\sqrt{2s_1-1} + \frac{(\hat{s}_3+1)g_3}{\sqrt{2s_2-1}}]
 \Phi_1^{\alpha(\hat{s}_3)\beta(\hat{s}_2)\delta} \zeta_\delta
\Omega_2^{\gamma(\hat{s}_1)}{}_{\alpha(\hat{s}_3)}
\Omega_{3,\beta(\hat{s}_2)\gamma(\hat{s}_1)} \\
 && + [ - g_2 \sqrt{2s_3-2} + \frac{\hat{s}_1g_3}{\sqrt{2s_2-1}} ]
\Phi_1^{\alpha(\hat{s}_3+1)\beta(\hat{s}_2)}
\Omega_2^{\gamma(\hat{s}_1-1)}{}_{\alpha(\hat{s}_3+1)}
\Omega_{3,\beta(\hat{s}_2)\gamma(\hat{s}_1-1)\delta} \zeta^\delta
\end{eqnarray*}
This gives:
\begin{eqnarray}
g_1 &=& \frac{\hat{s}_2}{(\hat{s}_3+1)} 
\sqrt{\frac{(2s_2-1)}{(2s_3-1)}} g_0 \nonumber \\
g_2 &=& - \frac{\hat{s}_1}{(\hat{s}_3+1)} 
\sqrt{\frac{(2s_1-1)}{(2s_3-1)}} g_0 \\
g_3 &=& - \frac{\sqrt{(2s_1-2)(2s_2-2)}}{(\hat{s}_3+1)} g_0 \nonumber
\end{eqnarray}
Now, calculating the variations with three fermions we obtain:
\begin{eqnarray*}
\Delta_2 &=& - \lambda^2 [ g_1 \sqrt{2s_1-1} +
\frac{g_3\hat{s}_2}{\sqrt{2s_3-2}}  ]
 \Phi_1^{\alpha(\hat{s}_3+1)\beta(\hat{s}_2-1)\delta} \zeta_\delta
\Phi_2^{\gamma(\hat{s}_1)}{}_{\alpha(\hat{s}_3+1)}
\Phi_{3,\beta(\hat{s}_2-1)\gamma(\hat{s}_1)}  \\
 && + \lambda^2 [ - g_2 \sqrt{2s_2-1} 
+ \frac{g_3\hat{s}_1}{\sqrt{2s_3-2}}  ]
\Phi_a^{\alpha(\hat{s}_3+1)\beta(\hat{s}_2)}
\Phi_2^{\gamma(\hat{s}_1-1)\delta}{}_{\alpha(\hat{s}_3+1)}
\zeta_\delta \Phi_{3,\beta(\hat{s}_2)\gamma(\hat{s}_1-1)} = 0
\end{eqnarray*}

\noindent
{\bf Three half-integer superspins} $(s_i,s_i-1/2)$ In this case we
consider
\begin{eqnarray}
{\cal L}_1 &=& g_0 \Omega_1^{\alpha(\hat{s}_3)\beta(\hat{s}_2)}
\Omega_2^{\gamma(\hat{s}_1)}{}_{\alpha(\hat{s}_3)}
\Omega_{3,\beta(\hat{s}_2)\gamma(\hat{s}_1)} \nonumber \\
 && + ig_1\lambda \Omega_1^{\alpha(\hat{s}_3)\beta(\hat{s}_2)}
\Psi_2^{\gamma(\hat{s}_1-1)}{}_{\alpha(\hat{s}_3)}
\Psi_{3,\beta(\hat{s}_2)\gamma(\hat{s}_1-1)} \nonumber \\
 && + ig_2\lambda \Psi_1^{\alpha(\hat{s}_3)\beta(\hat{s}_2-1)}
\Omega_2^{\gamma(\hat{s}_1)}{}_{\alpha(\hat{s}_3)}
\Psi_{3,\beta(\hat{s}_2-1)\gamma(\hat{s}_1)} \nonumber \\
 && + ig_3\lambda \Psi_1^{\alpha(\hat{s}_3-1)\beta(\hat{s}_2)}
\Psi_2^{\gamma(\hat{s}_1)}{}_{\alpha(\hat{s}_3-1)}
\Omega_{3,\beta(\hat{s}_2)\gamma(\hat{s}_1)},
\end{eqnarray}
where $\hat{s}_{1,2,3} \ge 1$. All calculations are quite similar to
the previous case, so let us provide only the answer:
\begin{eqnarray}
g_1 &=& \frac{\hat{s}_1}{\sqrt{(2s_2-2)(2s_3-2)}} g_0 \nonumber \\
g_2 &=& \frac{\hat{s}_2}{\sqrt{(2s_1-2)(2s_3-2)}} g_0 \\
g_3 &=& \frac{\hat{s}_3}{\sqrt{(2s_1-2)(2s_2-2)}} g_0 \nonumber
\end{eqnarray}

\section{Massive fields}

In this section we consider massive bosonic and fermionic higher spin
fields. We work in the gauge invariant frame-like multispinor
formalism and mostly follow \cite{BSZ17a}.

\subsection{Free boson}

For the gauge invariant description of massive boson we need a number
of one-forms ($\Omega^{\alpha(2k)}$, $f^{\alpha(2k)}$), 
$1 \le k \le s-1$, one-form $A$ and zero-forms ($B^{\alpha(2)}$, 
$\pi^{\alpha(2)}$,  $\varphi$). Free Lagrangian in $AdS_3$ background
has the form:
\begin{eqnarray}
{\cal L}_0 &=& \sum_{k=1} (-1)^{k+1} [ k \Omega_{\alpha(2k-1)\beta}
e^\beta{}_\gamma \Omega^{\alpha(2k-1)\gamma} + \Omega_{\alpha(2k)}
D f^{\alpha(2k)} ] \nonumber \\
 && + E B_{\alpha(2)} B^{\alpha(2)} - B_{\alpha(2)} e^{\alpha(2)}
D A - E \pi_{\alpha(2)} \pi^{\alpha(2)} + \pi_{\alpha(2)}
E^{\alpha(2)} D \varphi \nonumber \\
 && + \sum_{k=1} (-1)^{k+1} a_k [ - \frac{(k+2)}{k}
\Omega_{\alpha(2k)\beta(2)} e^{\beta(2)} f^{\alpha(2k)} + 
\Omega_{\alpha(2k)} e_{\beta(2)} f^{\alpha(2k)\beta(2)} ] \nonumber \\
 && + 2a_0 \Omega_{\alpha(2)} e^{\alpha(2)} A - a_0 
f_{\alpha\beta} E^\beta{}_\gamma B^{\alpha\gamma} + 2Ms
\pi_{\alpha(2)} E^{\alpha(2)} A \nonumber \\
 && + \sum_{k=1} (-1)^{k+1} b_k f_{\alpha(2k-1)\beta} 
e^\beta{}_\gamma f^{\alpha(2k-1)\gamma} + \frac{Msa_0}{2}
f_{\alpha(2)} E^{\alpha(2)} \varphi + \frac{3}{2}a_0{}^2 E \varphi^2
\label{gm}
\end{eqnarray}
Here the coefficients $(a,b)$ are determined by the requirement that
the Lagrangian must be gauge invariant and appear to be:
\begin{eqnarray}
a_k{}^2 &=& \frac{k(s+k+1)(s-k-1)}{2(k+1)(k+2)(2k+3)}
[M^2 - (k+1)^2\lambda^2], \nonumber \\
a_0{}^2 &=& \frac{(s+1)(s-1)}{3}[M^2 - \lambda^2], \\
b_k &=& \frac{M^2s^2}{4k(k+1)^2}, \qquad
M^2 = m^2 + (s-1)^2\lambda^2. \nonumber
\end{eqnarray}
Gauge transformations leaving this Lagrangian invariant look like:
\begin{eqnarray}
\delta \Omega^{\alpha(2k)} &=& D \eta^{\alpha(2k)} + \frac{(k+2)}{k}
a_k e_{\beta(2)} \eta^{\alpha(2k)\beta(2)} 
+ \frac{a_{k-1}}{k(2k-1)} e^{\alpha(2)} \eta^{\alpha(2k-2)} +
\frac{b_k}{k} e^\alpha{}_\beta \xi^{\alpha(2k-1)\beta}, \nonumber \\
\delta f^{\alpha(2k)} &=& D \xi^{\alpha(2k)} + a_k e_{\beta(2)}
\xi^{\alpha(2k)\beta(2)} + \frac{(k+1)a_{k-1}}{k(k-1)(2k-1)}
e^{\alpha(2)} \xi^{\alpha(2k-2)} + e^\alpha{}_\beta 
\eta^{\alpha(2k-1)\beta}, \nonumber \\
\delta \Omega^{\alpha(2)} &=& D \eta^{\alpha(2)} + 3_1
e_{\beta(2)} \eta^{\alpha(2)\beta(2)} + b_1 
e^\alpha{}_\beta \xi^{\alpha\beta}, \nonumber \\
\delta f^{\alpha(2)} &=& D \xi^{\alpha(2)} + e^\alpha{}_\beta
\eta^{\alpha\beta} + a_1 e_{\beta(2)} \xi^{\alpha(2)\beta(2)}
+ 2a_0 e^{\alpha(2)} \xi, \\
\delta B^{\alpha(2)} &=& 2a_0 \eta^{\alpha(2)}, \qquad
\delta A = D \xi + \frac{a_0}{4} e_{\alpha(2)}
\xi^{\alpha(2)}, \nonumber \\
\delta \pi^{\alpha(2)} &=& \frac{Msa_0}{2} \xi^{\alpha(2)}, \qquad
\delta \varphi = - 2Ms \xi. \nonumber
\end{eqnarray}

Unfortunately, in this general case it is impossible to make a
separation of variables similarly to the massless case. But this
becomes possible after the partial gauge fixing. Indeed, let us set
$\varphi = 0$ and solve its equation:
\begin{equation}
\varphi = 0 \quad \Rightarrow \quad
A = \frac{1}{2Ms} e_{\alpha(2)} \pi^{\alpha(2)}
\end{equation}
Resulting Lagrangian (after the rescaling $\pi \Rightarrow 
\frac{Ms}{2}\pi$) acquires the form:
\begin{eqnarray}
{\cal L}_0 &=& \sum_{k=1} (-1)^{k+1} 
[ k \Omega_{\alpha(2k-1)\beta} e^\beta{}_\gamma 
\Omega^{\alpha(2k-1)\gamma} + \Omega_{\alpha(2k)} D f^{\alpha(2k)}
+ \frac{M^2s^2}{4k(k+1)^2} f_{\alpha(2k-1)\beta} e^\beta{}_\gamma
f^{\alpha(2k-1)\gamma} \nonumber \\
 && \qquad + a_k ( - \frac{(k+2)}{k} \Omega_{\alpha(2k)\beta(2)}
e^{\beta(2)} f^{\alpha(2k)} + \Omega_{\alpha(2k)} e_{\beta(2)}
f^{\alpha(2k)\beta(2)} )] \\
 && + E B_{\alpha(2)} B^{\alpha(2)} - B_{\alpha\beta} 
E^\beta{}_\gamma D \pi^{\alpha\gamma} 
 - 2a_0 \Omega_{\alpha\beta} E^\beta{}_\gamma \pi^{\alpha\gamma}
 - a_0 f_{\alpha\beta} E^\beta{}_\gamma B^{\alpha\gamma} 
 + \frac{M^2s^2}{4} E \pi_{\alpha(2)} \pi^{\alpha(2)}. \nonumber
\end{eqnarray}
Now let us introduce:
\begin{eqnarray}
\Omega_\pm^{\alpha(2k)} &=& \Omega^{\alpha(2k)} \pm \frac{Ms}{2k(k+1)}
f^{\alpha(2k)}, \nonumber \\
 B_\pm^{\alpha(2)} &=& B^{\alpha(2)} \pm \frac{Ms}{2} \pi^{\alpha(2)}
\end{eqnarray}
and similarly for the gauge parameters. Then we obtain:
\begin{equation}
{\cal L}_0 = \frac{1}{2Ms} [ {\cal L}_+(\Omega_+,B_+) -
{\cal L}_-(\Omega_-,B_-) ],
\end{equation}
\begin{eqnarray}
{\cal L}_\pm &=& \sum_{k=1} (-1)^{k+1} [ k(k+1) 
\Omega_{\pm,\alpha(2k)} D \Omega_\pm^{\alpha(2k)} \pm
Msk \Omega_{\pm,\alpha(2k-1)\beta} e^\beta{}_\gamma
\Omega_\pm{}^{\alpha(2k-1)\gamma} \nonumber \\
 && \qquad \qquad - 2(k+1)(k+2) a_k \Omega_{\pm.\alpha(2k)\beta(2)}
e^{\beta(2)} \Omega_\pm^{\alpha(2k)} ] \nonumber \\
 && - B_{\pm,\alpha\beta} E^\beta{}_\gamma D B_\pm^{\alpha\gamma}
  \pm Ms E B_{\pm,\alpha(2)} B_\pm^{\alpha(2)}
 - 4a_0 \Omega_{\pm,\alpha\beta} E^\beta{}_\gamma
B_\pm^{\alpha\gamma}. 
\end{eqnarray}
Each Lagrangian ${\cal L}_\pm$ is invariant under its own set of gauge
transformations:
\begin{eqnarray}
\delta \Omega_\pm^{\alpha(2k)} &=& D \eta_\pm^{\alpha(2k)} +
\frac{(k+2)}{k}a_k e_{\beta(2)} \eta_\pm^{\alpha(2k)\beta(2)} +
\frac{a_{k-1}}{k(2k-1)} e^{\alpha(2)} \eta_\pm{}^{\alpha(2k-2)}
\nonumber \\
 && \pm \frac{Ms}{2k(k+1)} e^\alpha{}_\beta 
\eta_\pm^{\alpha(2k-1)\beta}, \nonumber \\
\delta \Omega_\pm^{\alpha(2)} &=& D \eta_\pm^{\alpha(2)} + 3a_1
e_{\beta(2)} \eta_\pm^{\alpha(2)\beta(2)} \pm \frac{Ms}{4} 
e^\alpha{}_\beta \eta_\pm^{\alpha\beta}, \\
\delta B_\pm^{\alpha(2)} &=& 2a_0 \eta_\pm^{\alpha(2)}. \nonumber
\end{eqnarray}
As in the massless case, from now on we work only with $+$ components,
omitting the $+$ sign. 

Now we try to construct a complete set of gauge invariant objects (we
still call all them curvatures though now they are two and one-forms).
For the one-forms the structure of the gauge invariant two-forms can
be easily read from the structure of gauge transformations:
\begin{eqnarray}
{\cal R}^{\alpha(2k)} &=& D \Omega^{\alpha(2k)} + \frac{(k+2)}{k}a_k
e_{\beta(2)} \Omega^{\alpha(2k)\beta(2)} + \frac{a_{k-1}}{k(2k-1)}
e^{\alpha(2)} \Omega^{\alpha(2k-2)} \nonumber \\
 && + \frac{Ms}{2k(k+1)} e^\alpha{}_\beta 
\Omega^{\alpha(2k-1)\beta}, \\
{\cal R}^{\alpha(2)} &=& D \Omega^{\alpha(2)} + 3a_1 e_{\beta(2)}
\Omega^{\alpha(2)\beta(2)} + \frac{Ms}{4} e^\alpha{}_\beta
\Omega^{\alpha\beta} + c_0 E^\alpha{}_\beta B^{\alpha\beta}. \nonumber
\end{eqnarray}
But to construct a gauge invariant one-form for $B^{\alpha(2)}$ we
have to introduce an extra zero-form $B^{\alpha(4)}$:
\begin{equation}
{\cal B}^{\alpha(2)} = D B^{\alpha(2)} - 2a_0 \Omega^{\alpha(2)}
+ 3a_1 e^\alpha{}_\beta B^{\alpha\beta} + \frac{Ms}{4} e_{\beta(2)}
B^{\alpha(2)\beta(2)},
\end{equation}
where we postulate
\begin{equation}
\delta B^{\alpha(4)} = 2a_0 \eta^{\alpha(4)}
\end{equation}
Then to construct a curvature for $B^{\alpha(4)}$ we need
$B^{\alpha(6)}$ and so on. The procedure stops with the set of extra
zero-forms with $2 \le k \le s-1$
\begin{equation}
\delta B^{\alpha(2k)} = 2a_0 \eta^{\alpha(2k)},
\end{equation}
with their curvatures (one-forms) being:
\begin{eqnarray}
{\cal B}^{\alpha(2k)} &=& D B^{\alpha(2k)} - 2a_0 \Omega^{\alpha(2k)}
+ \frac{Ms}{2k(k+1)} e^\alpha{}_\beta B^{\alpha(2k-1)\beta} \nonumber
\\
 && + \frac{(k+2)}{k}a_k e_{\beta(2)} B^{\alpha(2k)\beta(2)} + 
\frac{a_{k-1}}{k(2k-1)} e^{\alpha(2)} B^{\alpha(2k-2)} 
\end{eqnarray}
With the help of all these gauge invariant curvatures, one can rewrite
the Lagrangian in the explicitly gauge invariant form:
\begin{equation}
{\cal L}_0 = \sum_{k=1}^{s-1} (-1)^k \frac{k(k+1)}{2a_0}
{\cal R}_{\alpha(2k)} {\cal B}^{\alpha(2k)},
\end{equation}
where coefficients are determined by the so-called extra field
decoupling condition:
$$
\frac{\delta {\cal L}_0}{\delta B^{\alpha(2k)}} = 0, \qquad k > 1.
$$

A large number of fields involved in the description of free field
make the investigation of their interactions very cumbersome. To
simplify investigations, we use the procedure of (almost) maximal
gauge fixing. In more details, let us use $\eta^{\alpha(2)}$
transformations to set $B^{\alpha(2)}=0$ and set to zero its gauge
invariant one-form:
\begin{equation}
B^{\alpha(2)} = 0 \quad \Rightarrow \quad
\Omega^{\alpha(2)} = \frac{3a_1}{2a_0} e_{\beta(2)} 
B^{\alpha(2)\beta(2)},
\end{equation}
then it easy to check that
\begin{eqnarray}
{\cal R}^{\alpha(2)} &=& - \frac{3a_1}{2a_0} [ e_{\beta(2)} D
B^{\alpha(2)\beta(2)} - 2a_0 e_{\beta(2)} \Omega^{\alpha(2)\beta(2)} +
Ms E_{\beta(2)} B^{\alpha(2)\beta(2)} ] \nonumber \\
 &=& - \frac{3a_1}{2a_0} e_{\beta(2)} {\cal B}^{\alpha(2)\beta(2)},
\end{eqnarray}
so that ${\cal R}^{\alpha(2)}$ is not an independent object and can be
omitted. Proceeding in this way, at the last step we set
\begin{equation}
B^{\alpha(2s-4)} = 0 \quad \Rightarrow \quad
\Omega^{\alpha(2s-4)} = \frac{sa_{s-2}}{2(s-2)a_0} e_{\beta(2)}
B^{\alpha(2s-4)\beta(2)}.
\end{equation}
Here also we find that
\begin{equation}
{\cal R}^{\alpha(2s-4)} = \frac{sa_{s-2}}{2(s-2)a_0} e_{\beta(2)}
{\cal B}^{\alpha(2s-4)\beta(2)} 
\end{equation}
is not independent any more.

As a result, we obtain really minimal Lagrangian (rescaling
$B^{\alpha(2s-2)}$ for convenience):
\begin{eqnarray}
\frac{(-1)^s}{s(s-1)}{\cal L} &=& \Omega_{\alpha(2s-2)} D
\Omega^{\alpha(2s-2)} + M \Omega_{\alpha(2s-3)\beta} e^\beta{}_\gamma
\Omega^{\alpha(2s-3)\gamma} \nonumber \\
 && - B_{\alpha(2s-3)\beta} E^\beta{}_\gamma D
B^{\alpha(2s-3)\gamma}  + \frac{Ms}{(s-1)} E B_{\alpha(2s-2)}
B^{\alpha(2k-2)} \nonumber \\
 && - 2\tilde{m} \Omega_{\alpha(2s-3)\beta} E^\beta{}_\gamma
B^{\alpha(2s-3)\gamma}, 
\end{eqnarray}
where
$$
\tilde{m} = \sqrt{\frac{2}{(s-1)}} m.
$$
This Lagrangian follows the general pattern for the gauge invariant
description for the massive fields. Indeed, it is invariant under the
only remaining gauge transformations:
\begin{eqnarray}
\delta \Omega^{\alpha(2s-2)} &=& D \eta^{\alpha(2s-2)} 
+ \frac{M}{2(s-1)} e^\alpha{}_\beta \eta^{\alpha(2s-3)\beta},
\nonumber \\
\delta B^{\alpha(2s-2)} &=& \tilde{m} \eta^{\alpha(2s-2)}.
\end{eqnarray}
Moreover, we still have a couple of gauge invariant curvatures:
\begin{eqnarray}
{\cal R}^{\alpha(2s-2)} &=& D \Omega^{\alpha(2s-2)} + \frac{M}{2(s-1)}
e^\alpha{}_\beta \Omega^{\alpha(2s-3)\beta} - \frac{\tilde{m}}{2(s-1)}
E^\alpha{}_\beta B^{\alpha(2s-3)\beta}, \nonumber \\
{\cal B}^{\alpha(2s-2)} &=& D B^{\alpha(2s-2)} - \tilde{m}
\Omega^{\alpha(2s-2)} + \frac{M}{2(s-1)} e^\alpha{}_\beta
B^{\alpha(2s-3)\beta}. 
\end{eqnarray}
They satisfy the following differential identities:
\begin{eqnarray}
D {\cal R}^{\alpha(2s-2)} &=& - \frac{M}{2(s-1)} e^\alpha{}_\beta
{\cal R}^{\alpha(2s-3)\beta} - \frac{\tilde{m}}{2(s-1)}
E^\alpha{}_\beta {\cal B}^{\alpha(2s-3)\beta}, \nonumber \\
D {\cal B}^{\alpha(2s-2)} &=& - \tilde{m} {\cal R}^{\alpha(2s-2)}
- \frac{M}{2(s-1)} e^\alpha{}_\beta {\cal B}^{\alpha(2s-3)\beta}. 
\end{eqnarray}
Naturally, these curvatures appear in the variation of the Lagrangian
under the arbitrary variations of $\Omega$ and $B$ fields:
\begin{equation}
\delta {\cal L} \sim {\cal R}_{\alpha(2s-2)} \delta
\Omega^{\alpha(2s-2)} + {\cal B}_{\alpha(2s-3)\beta}
E^\beta{}_\gamma \delta B^{\alpha(2s-3)\gamma}.
\end{equation}
At last, but not least, the Lagrangian can be nicely rewritten in the
explicitly gauge invariant form:
\begin{equation}
{\cal L} \sim \frac{1}{2\tilde{m}} {\cal R}_{\alpha(2s-2)}
{\cal B}^{\alpha(2s-2)}.
\end{equation}

With this Lagrangian formulation we can associate a self-consistent
set
of unfolded equations (compare \cite{BPSS15,Zin15,BSZ16}):
\begin{eqnarray}
0 &=& D \Omega^{\alpha(2s-2)} + \frac{M}{2(s-1)} e^\alpha{}_\beta
\Omega^{\alpha(2s-3)\beta} - \frac{\tilde{m}}{2(s-1)} E^\alpha{}_\beta
B^{\alpha(2s-3)\beta}, \nonumber \\
 0 &=& D B^{\alpha(2s-2)} - \tilde{m} \Omega^{\alpha(2s-2)} 
+ \frac{M}{2(s-1)} e^\alpha{}_\beta B^{\alpha(2s-3)\beta} +
e_{\beta(2)} W^{\alpha(2s-2)\beta(2)}, \\
0 &=& D W^{\alpha(2k)} + e_{\beta(2)} W^{\alpha(2k)\beta(2)}
+ \alpha_k e^\alpha{}_\beta W^{\alpha(2k-1)\beta} +
\beta_k e^{\alpha(2)} W^{\alpha(2k-2)}, \nonumber
\end{eqnarray}
where $W^{\alpha(2k)}$, $k \ge s$ is an infinite set of gauge
invariant zero-forms and
\begin{equation}
\alpha_k = \frac{Ms}{2k(k+1)}, \qquad
\beta_k = - \frac{(k^2-s^2)}{2(4k^2-1)} [\frac{M^2}{k^2} - \lambda^2].
\end{equation}
Note, that the equations for $W^{\alpha(2k)}$ have exactly the same
form as in the general case \cite{Zin15,BSZ16} and this serves as an
additional conformation that after all these gauge fixing we still
have one physical degree of freedom.

\subsection{Free fermion}

For the gauge invariant description of massive spin-$(s+1/2)$ fermion
we need a set of one-forms $\Phi^{\alpha(2k+1)}$, $(0 \le k \le s-1)$
and zero-form $\phi^\alpha$. Free Lagrangian looks like:
\begin{eqnarray}
\frac{1}{i} {\cal L}_0 &=& \sum_{k=0}^{s-1} (-1)^{k+1}
[ \frac{1}{2} \Phi_{\alpha(2k+1)} D \Phi^{\alpha(2k+1)}]
+ \frac{1}{2} \phi_\alpha E^\alpha{}_\beta D \phi^\beta \nonumber \\
 && + \sum_{k=1}^{s-1} (-1)^{k+1} c_k \Phi_{\alpha(2k-1)\beta(2)}
e^{\beta(2)} \Phi^{\alpha(2k-1)} + c_0 \Phi_\alpha E^\alpha{}_\beta
\phi^\beta \nonumber \\
 && + \sum_{k=0}^{s-1} (-1)^{k+1} \frac{d_k}{2} \Phi_{\alpha(2k)\beta}
e^\beta{}_\gamma \Phi^{\alpha(2k)\gamma} - \frac{3d_0}{2} E
\phi_\alpha \phi^\alpha,
\end{eqnarray}
where
\begin{eqnarray}
d_k &=& \frac{(2s+1)}{(2k+3)}M, \qquad
M^2 = m^2 + (s-\frac{1}{2})^2 \lambda^2, \nonumber \\
c_k{}^2 &=& \frac{(s+k+1)(s-k)}{2(k+1)(2k+1)}
[M^2 - (2k+1)^2\frac{\lambda^2}{4} ], \\
c_0{}^2 &=& 2s(s+1)[M^2 - \frac{\lambda^2}{4}]. \nonumber
\end{eqnarray}
This Lagrangian is invariant under the following gauge
transformations:
\begin{eqnarray}
\delta_0 \Phi^{\alpha(2k+1)} &=& D \zeta^{\alpha(2k+1)} +
\frac{d_k}{(2k+1)} e^\alpha{}_\beta \zeta^{\alpha(2k)\beta} 
 + \frac{c_k}{k(2k+1)} e^{\alpha(2)} \zeta^{\alpha(2k-1)} \nonumber \\
 && + c_{k+1} e_{\beta(2)} \zeta^{\alpha(2k+1)\beta(2)}, \\
\delta_0 \phi^\alpha &=& c_0 \zeta^\alpha. \nonumber
\end{eqnarray}
As in the bosonic case, to construct a complete set of the gauge
invariant curvatures:
\begin{eqnarray}
{\cal F}^{\alpha(2k+1)} &=& D \Phi^{\alpha(2k+1)} +
\frac{d_k}{(2k+1)} e^\alpha{}_\beta \Phi^{\alpha(2k)\beta} \nonumber 
\\
 && + \frac{c_k}{k(2k+1)} e^{\alpha(2)} \Phi^{\alpha(2k-1)} 
+ c_{k+1} e_{\beta(2)} \Phi^{\alpha(2k+1)\beta(2)}, \nonumber \\
{\cal F}^\alpha &=& D \Phi^\alpha + d_0 e^\alpha{}_\beta \Phi^\beta
+ c_1 e_{\beta(2)} \Phi^{\alpha\beta(2)} - c_0 E^\alpha{}_\beta
\phi^\beta, \nonumber \\
{\cal C}^\alpha &=& D \phi^\alpha - c_0 \Phi^\alpha + d_0
e^\alpha{}_\beta \phi^\beta + c_1 e_{\beta(2)} \phi^{\alpha\beta(2)}, 
\\
{\cal C}^{\alpha(2k+1)} &=& D \phi^{\alpha(2k+1)} - c_0
\Phi^{\alpha(2k+1)} + \frac{d_k}{(2k+1)} e^\alpha{}_\beta
\phi^{\alpha(2k)\beta}, \nonumber \\
 && + \frac{c_k}{k(2k+1)} e^{\alpha(2)} \phi^{\alpha(2k-1)}
+ c_{k+1} e_{\beta(2)} \phi^{\alpha(2k+1)\beta(2)}, \nonumber
\end{eqnarray}
we have to introduce a number of extra zero-forms 
$\phi^{\alpha(2k+1)}$, $1 \le k \le s-1$, where
\begin{equation}
\delta \phi^{\alpha(2k+1)} = c_0 \zeta^{\alpha(2k+1)}.
\end{equation}
Such Lagrangian can also be rewritten in terms of the curvatures:
\begin{equation}
{\cal L}_0 = \sum_{k=0}^{s-1} (-1)^k \frac{i}{2c_0} 
{\cal F}_{\alpha(2k+1)} {\cal C}^{\alpha(2k+1)},
\end{equation}
where coefficients are again determined by the extra field decoupling
conditions:
\begin{equation}
\frac{\delta {\cal L}_0}{\delta \phi^{\alpha(2k+1)}} = 0, \qquad 
k > 0.
\end{equation}

As in the bosonic case, we now apply the (almost) maximal gauge
fixing. We begin by setting
\begin{equation}
\phi^\alpha = 0 \quad \Rightarrow \quad
\Phi^\alpha = \frac{c_1}{c_0} e_{\beta(2)} \phi^{\alpha\beta(2)}
\end{equation}
and checking that
\begin{equation}
{\cal F}^\alpha = - \frac{c_1}{c_0} e_{\beta(2)} 
{\cal C}^{\alpha\beta(2)}.
\end{equation}
Proceeding in this way, at the last step we set:
\begin{equation}
\phi^{\alpha(2s-3)} = 0 \quad \Rightarrow \quad
\Phi^{\alpha(2s-3)} = \frac{c_{s-1}}{c_0} e_{\beta(2)}
\phi^{\alpha(2s-3)\beta(2)} 
\end{equation}
and check that
\begin{equation}
{\cal F}^{\alpha(2s-3)} = - \frac{c_{s-1}}{c_0} e_{\beta(2)}
{\cal C}^{\alpha(2s-3)\beta(2)}.
\end{equation}
Thus we obtain the minimal Lagrangian (rescaling 
$\phi^{\alpha(2s-1)}$ for simplicity):
\begin{eqnarray}
\frac{(-1)^s}{i} {\cal L} &=& \frac{1}{2} \Phi_{\alpha(2s-1)} D
\Phi^{\alpha(2s-1)}  - 2 \phi_{\alpha(2s-2)\beta} E^\beta{}_\gamma D
\phi^{\alpha(2s-2)\gamma} \nonumber \\
 && - 4\tilde{m} \Phi_{\alpha(2s-2)\beta} E^\beta{}_\gamma
\phi^{\alpha(2s-2)\gamma} \nonumber \\
 && + \frac{M}{2} \Phi_{\alpha(2s-2)\beta} e^\beta{}_\gamma
\Phi^{\alpha(2s-2)\gamma} - \frac{2(2s+1)}{(2s-1)}M E
\phi_{\alpha(2s-1)} \phi^{\alpha(2s-1)},
\end{eqnarray}
where
\begin{equation}
\tilde{m} = \sqrt{\frac{m}{(2s-1)}} .
\end{equation}
This Lagrangian is still invariant under the only remaining gauge
transformations:
\begin{eqnarray}
\delta_0 \Phi^{\alpha(2s-1)} &=& D \zeta^{\alpha(2s-1)} 
+ \frac{M}{(2s-1)} e^\alpha{}_\beta \zeta^{\alpha(2s-2)\beta},
\nonumber \\
\delta_0 \phi^{\alpha(2s-1)} &=& \tilde{m} \zeta^{\alpha(2s-1)}
\end{eqnarray}
Also, we still have a couple of gauge invariant curvatures:
\begin{eqnarray}
{\cal F}^{\alpha(2s-1)} &=& D \Phi^{\alpha(2s-1)} +
\frac{M}{(2s-1)} e^\alpha{}_\beta \Phi^{\alpha(2s-2)\beta}
- \frac{4c_{s-1}}{(2s-1)} E^\alpha{}_\beta \phi^{\alpha(2s-2)\beta},
\nonumber \\
{\cal C}^{\alpha(2s-1)} &=& D \phi^{\alpha(2s-1)} - c_{s-1}
\Phi^{\alpha(2s-1)} + \frac{M}{(2s-1)} e^\alpha{}_\beta,
\phi^{\alpha(2s-2)\beta} 
\end{eqnarray}
satisfying the following differential identities:
\begin{eqnarray}
D {\cal F}^{\alpha(2s-1)} &=& - \frac{M}{(2s-1)} e^\alpha{}_\beta
{\cal F}^{\alpha(2s-2)\beta} - \frac{4c_{s-1}}{(2s-1)}
E^\alpha{}_\beta {\cal C}^{\alpha(2s-2)\beta}, \nonumber \\
D {\cal C}^{\alpha(2s-1)} &=& - c_{s-1} {\cal F}^{\alpha(2s-1)}
- \frac{M}{(2s-1)} e^\alpha{}_\beta {\cal C}^{\alpha(2s-2)\beta}.
\end{eqnarray}
Variation of the Lagrangian under the arbitrary variations of $\Phi$
and $\phi$ has the form:
\begin{equation}
\delta {\cal L} \sim {\cal F}_{\alpha(2s-1)} \delta
\Phi^{\alpha(2s-1)} + 4 {\cal C}_{\alpha(2s-2)\beta} E^\beta{}_\gamma
\delta \phi^{\alpha(2s-2)\gamma},
\end{equation}
while the Lagrangian can be rewritten simply as
\begin{equation}
{\cal L} \sim \frac{1}{2\tilde{m}} {\cal F}_{\alpha(2s-1)}
{\cal C}^{\alpha(2s-1)}. 
\end{equation}

Here there also exists a set of self-consistent unfolded equations
(compare \cite{BSZ16}):
\begin{eqnarray}
0 &=& D \Phi^{\alpha(2s-1)} + \frac{M}{(2s-1)} e^\alpha{}_\beta
\Phi^{\alpha(2s-2)\beta} - \frac{4c_{s-1}}{(2s-1)} E^\alpha{}_\beta
\phi^{\alpha(2s-2)\beta}, \nonumber \\
0 &=& D \phi^{\alpha(2s-1)} - c_{s-1}
\Phi^{\alpha(2s-1)} + \frac{M}{(2s-1)} e^\alpha{}_\beta
\phi^{\alpha(2s-2)\beta} + e_{\beta(2)} V^{\alpha(2s-1)\beta(2)}, \\
0 &=& D V^{\alpha(2k+1)} + e_{\beta(2)} V^{\alpha(2k+1)\beta(2)}
+ \alpha_k e^\alpha{}_\beta V^{\alpha(2k)\beta} + \beta_k
e^{\alpha(2)} V^{\alpha(2k-1)}, \nonumber
\end{eqnarray}
where $V^{\alpha(2k+1)}$, $k \ge s$ is an infinite set of gauge
invariant fermionic zero-forms, while
\begin{equation}
\alpha_k = \frac{(2s+1)M}{(2k+1)(2k+3)}, \qquad
\beta_k = - \frac{(k-s)(k+s+1)}{2k(k+1)}
[\frac{M^2}{(2k+1)^2} - \frac{\lambda^2}{4}].
\end{equation}
Let us stress once again, that the equations for $V$ have the same
form as in the general formalism without any gauge fixing.

\subsection{Two massive and one massless fields}

In this subsection we consider possible cubic vertices for two massive
and one massless fields. Let us begin with the purely bosonic case:
two massive bosons with spins $s_{1,2}$ and one massless boson with
spin $s_3$. Any consistent non-abelian corrections to the gauge
transformations must be related with consistent deformations for all
gauge invariant curvatures \cite{Zin16} (schematically):
\begin{eqnarray*}
\Delta {\cal R}_1 &\sim& \Omega_2 \Omega_3 + e B_2 \Omega_3, \qquad
\Delta {\cal B}_1 \sim B_2 \Omega_3, \\
\Delta {\cal R}_2 &\sim& \Omega_1 \Omega_3 + e B_1 \Omega_3, \qquad
\Delta {\cal B}_2 \sim B_2 \Omega_3.
\end{eqnarray*}
The main consistency requirement here is that the deformed curvatures
$\hat{\cal R}_{1,2} = {\cal R}_{1,2} + \Delta {\cal R}_{1,2}$,
$\hat{\cal B}_{1,2} = {\cal B}_{1,2} + \Delta {\cal B}_{1,2}$
transform covariantly:
\begin{eqnarray*}
\delta \hat{\cal R}_1 &\sim& {\cal R}_2 \eta_3 
+ e {\cal B}_2 \eta_3, \qquad
\delta \hat{\cal B}_1 \sim {\cal B}_2 \eta_3, \\
\delta \hat{\cal R}_2 &\sim& {\cal R}_1 \eta_3 
+ e {\cal B}_1 \eta_3, \qquad
\delta \hat{\cal B}_2 \sim {\cal B}_1 \eta_3.
\end{eqnarray*}
Due to the presence of zero-form fields, there exists a pair of
possible field redefinitions:
$$
\Omega_1 \Rightarrow \Omega_1 + \kappa_1 B_2 \Omega_3, \qquad
\Omega_2 \Rightarrow \Omega_2 + \kappa_2 B_1 \Omega_3.
$$
It appears that using these field redefinitions one can completely
remove all possible non-abelian corrections. Taking into account that
in three dimensions it is not possible to construct any trivially
gauge invariant vertex (simply because it must be four-form as a
minimum), the only remaining possibility is an abelian vertex:
\begin{equation}
{\cal L}_1 = g {\cal B}_1^{\alpha(\hat{s}_3)\beta(\hat{s}_2)}
{\cal B}_2^{\gamma(\hat{s}_1)}{}_{\alpha(\hat{s}_3)}
\Omega_{3,\beta(\hat{s}_2)\gamma(\hat{s}_1)}.
\end{equation}
Here $\hat{s}_i$ are the same as in the massless case and it means
that three spins $s_{1,2,3}$ also must satisfy a strict triangular
inequality. Consider variation of the vertex under the gauge
transformations of the massless field:
\begin{eqnarray*}
\frac{1}{g} \delta {\cal L}_1 &=& 
{\cal B}_1^{\alpha(\hat{s}_3)\beta(\hat{s}_2)}
{\cal B}_2^{\gamma(\hat{s}_1)}{}_{\alpha(\hat{s}_3)}
[ D \eta_{\beta(\hat{s}_2)\gamma(\hat{s}_1} - \frac{\lambda}{2}
(e_\beta{}^\delta \eta_{\beta(\hat{s}_2-1)\delta\gamma(\hat{s}_1)} +
e_\gamma{}^\delta \eta_{\beta(\hat{s}_2)\gamma(\hat{s}_1-1)\delta})] 
\\
 &&  \\
 &=& \tilde{m}_1 {\cal R}_1^{\alpha(\hat{s}_3)\beta(\hat{s}_2)}
{\cal B}_2^{\gamma(\hat{s}_1)}{}_{\alpha(\hat{s}_3)}
\eta_{\beta(\hat{s}_2)\gamma(\hat{s}_1)} 
 - \tilde{m}_2 {\cal B}_1^{\alpha(\hat{s}_3)\beta(\hat{s}_2)}
{\cal R}_2^{\gamma(\hat{s}_1)}{}_{\alpha(\hat{s}_3)}
\eta_{\beta(\hat{s}_2)\gamma(\hat{s}_1)} \\
 &&  \\
 && + \frac{M_1}{(2s_1-2)} (e^\alpha{}_\delta 
{\cal B}_1^{\alpha(\hat{s}_3-1)\delta\beta(\hat{s}_2)}
+ e^\beta{}_\delta 
{\cal B}_1^{\alpha(\hat{s}_3)\beta(\hat{s}_2-1)\delta})
{\cal B}_2^{\gamma(\hat{s}_1)}{}_{\alpha(\hat{s}_3)}
\eta_{\beta(\hat{s}_2)\gamma(\hat{s}_1)} \\
 && - \frac{M_2}{(2s_2-2)} 
{\cal B}_1^{\alpha(\hat{s}_3)\beta(\hat{s}_2)} (e^\gamma{}_\delta
{\cal B}_2^{\gamma(\hat{s}_1-1)\delta}{}_{\alpha(\hat{s}_3)} -
e_\alpha{}^\delta 
{\cal B}_2^{\gamma(\hat{s}_1)}{}_{\alpha(\hat{s}_3-1)\delta})
\eta_{\beta(\hat{s}_2)\gamma(\hat{s}_1)} \\
 && - \frac{\lambda}{2} {\cal B}_1^{\alpha(\hat{s}_3)\beta(\hat{s}_2)}
{\cal B}_2^{\gamma(\hat{s}_1)}{}_{\alpha(\hat{s}_3)}
(e_\beta{}^\delta \eta_{\beta(\hat{s}_2-1)\delta\gamma(\hat{s}_1)} +
e_\gamma{}^\delta \eta_{\beta(\hat{s}_2)\gamma(\hat{s}_1-1)\delta}).
\end{eqnarray*}
Terms with the ${\cal R}_{1,2}$ vanish on-shell or, if one likes, can
be compensated by the abelian corrections to the gauge
transformations:
\begin{eqnarray}
\delta \Omega_1^{\alpha(2s_1-2)} &\sim& 
{\cal B}_2^{\alpha(\hat{s}_3)\gamma(\hat{s}_1)}
\eta^{\alpha(\hat{s}_2)}{}_{\gamma(\hat{s}_1)}, \nonumber \\
\delta \Omega_2^{\alpha(2s_1-2)} &\sim& 
{\cal B}_1^{\alpha(\hat{s}_3)\gamma(\hat{s}_2)}
\eta^{\alpha(\hat{s}_1)}{}_{\gamma(\hat{s}_2)},
\end{eqnarray}
which correspond to abelian deformations for curvatures:
\begin{eqnarray}
\Delta {\cal R}_1^{\alpha(2s_1-2)} &\sim& 
{\cal B}_2^{\alpha(\hat{s}_3)\gamma(\hat{s}_1)}
\Omega_3^{\alpha(\hat{s}_2)}{}_{\gamma(\hat{s}_1)}, \nonumber \\
\Delta {\cal R}_2^{\alpha(2s_1-2)} &\sim& 
{\cal B}_1^{\alpha(\hat{s}_3)\gamma(\hat{s}_2)}
\Omega_3^{\alpha(\hat{s}_1)}{}_{\gamma(\hat{s}_2)}.
\end{eqnarray}
As for the terms quadratic in ${\cal B}$, we use the fact that 
on-shell
$$
{\cal B}^{\alpha(2s-2)} \approx e_{\beta(2)} W^{\alpha(2s-2)\beta(2)}.
$$
Then these terms are reduced to
\begin{eqnarray}
\Delta &\sim& [\frac{\hat{s}_2+\hat{s}_3}{(2s_1-2)}M_1
- \frac{\hat{s}_1+\hat{s}_3}{(2s_2-2)} M_2 
+ (\hat{s}_1-\hat{s}_2) \frac{\lambda}{2}] 
 W_1^{\alpha(\hat{s}_3+2)\beta(\hat{s}_2)}
W_2^{\gamma(\hat{s}_1)}{}_{\alpha(\hat{s}_3+2)}
\eta_{\beta(\hat{s}_2)\gamma(\hat{s}_1)} \nonumber \\
 &=& [ M_1 - M_2 - (s_1-s_2)\lambda ]
W_1^{\alpha(\hat{s}_3+2)\beta(\hat{s}_2)}
W_2^{\gamma(\hat{s}_1)}{}_{\alpha(\hat{s}_3+2)}
\eta_{\beta(\hat{s}_2)\gamma(\hat{s}_1)}.
\end{eqnarray}
Thus  the gauge invariance imposes the relation on the masses:
\begin{equation}
M_1 - M_2 = (s_1-s_2)\lambda. \label{gen_rel}
\end{equation}
In the same way one can construct cubic vertices for two other cases:
two massive fermions and one massless boson
\begin{equation}
{\cal L}_1 = g {\cal C}_1^{\alpha(\hat{s}_3+1)\beta(\hat{s}_2)}
{\cal C}_2^{\gamma(\hat{s}_1)}{}_{\alpha(\hat{s}_3+1)}
\Omega_{3,\beta(\hat{s}_2)\gamma(\hat{s}_1)}
\end{equation}
or massive boson and massive and massless fermions
\begin{equation}
{\cal L}_1 = g {\cal B}_1^{\alpha(\hat{s}_3)\beta(\hat{s}_2)}
{\cal C}_2^{\gamma(\hat{s}_1+1)}{}_{\alpha(\hat{s}_3)}
\Phi_{3,\beta(\hat{s}_2)\gamma(\hat{s}_1+1)}
\end{equation}
with the same relations on the masses.

Note, that for the first time a simple example of such mass relation
appeared in our construction of unfolded formulation for massive
higher spin supermultiplets \cite{BSZ16}. There we consider
deformations of the unfolded equations for massive boson $(M_1,s_1)$
and massive fermion $(M_2,s_2)$ in the presence of a background
massless spin-3/2 field:
\begin{equation}
0 = D \Phi^\alpha + \frac{\lambda}{2} e^\alpha{}_\beta \Phi^\beta.
\end{equation}
For the gauge invariant part of the unfolded equations such ansatz has
the form:
\begin{eqnarray}
0 &=& D W^{\alpha(2k)} + e_{\beta(2)} W^{\alpha(2k)\beta(2)}
+ \alpha_k e^\alpha{}_\beta W^{\alpha(2k-1)\beta} +
\beta_k e^{\alpha(2)} W^{\alpha(2k-2)} \nonumber \\
 && + f_{k,0} \Phi_\beta V^{\alpha(2k)\beta}
 + f_{k,1} \Phi^\alpha V^{\alpha(2k-1)}, \\
0 &=& D V^{\alpha(2k+1)} + e_{\beta(2)} V^{\alpha(2k+1)\beta(2)}
+ \gamma_k e^\alpha{}_\beta V^{\alpha(2k)\beta} + \delta_k
e^{\alpha(2)} V^{\alpha(2k-1)} \nonumber \\
 && + g_{k,0} \Phi_\beta W^{\alpha(2k+1)\beta} 
 + g_{k,1} \Phi^\alpha W^{\alpha(2k)}. 
\end{eqnarray}
Self-consistency of these deformations requires
\begin{equation}
s_2 = s_1 + \Delta, \qquad M_2 = M_1 + \Delta \lambda, \qquad
\Delta = \pm \frac{1}{2}, \label{rel}
\end{equation}
the solution being:
\begin{eqnarray}
f_{k,0} = f_0, \qquad
f_{k,1} = \mp \frac{(k \mp s_1)}{2k(2k+1)}[M_1 \mp k\lambda]f_0, \\
g_{k,0} = g_0, \qquad
g_{k,1} = \pm \frac{(k \pm s_1+1)}{2(k+1)(2k+1)} [M_1 \pm
(k+1)\lambda]g_0.
\end{eqnarray}
This construction can be easily generalized to the massless fermion
with arbitrary spin $s$ (see Appendix A). We have found a complete
solution for the case $s = \frac{5}{2}$ and obtain the same relations
as in (\ref{rel}) but with 
$\Delta = \pm \frac{1}{2}, \pm \frac{3}{2}$. 

It is instructive to compare our results with what happens in the
Prokushkin-Vasiliev theory \cite{PV98,PSV99}, describing interaction
of higher spin massless fields with massive spin 0 and spin 1/2. The
unfolded equations for these massive fields \cite{Vas94} in our
current notation look like:
\begin{equation}
0 = D W^{\alpha(2k)} + e_{\beta(2)} W^{\alpha(2k)\beta(2)}
+ \beta_k e^{\alpha(2)} W^{\alpha(2k-2)}
\end{equation}
$$
\beta_k = - \frac{1}{2(4k^2-1)}[m_0^2 - (k^2-1)\lambda^2]
$$
\begin{equation}
0 = D V^{\alpha(2k+1)} + e_{\beta(2)} V^{\alpha(2k+1)\beta(2)}
+ \alpha_k e^\alpha{}_\beta V^{\alpha(2k)\beta} + \beta_k
e^{\alpha(2)} V^{\alpha(2k-1)}
\end{equation}
$$
\alpha_k = \frac{m}{(2k+1)(2k+3)}, \qquad
\beta_k = - \frac{1}{2(2k+1)^2}[m^2 - (2k+1)^2\frac{\lambda^2}{4}]
$$
Trying to solve consistency relations given in appendix A iteratively
by $l$ already at $l=2$ we have found that the solution exists only
when
\begin{equation}
m_0{}^2 = m^2 - m\lambda - \frac{3}{4}\lambda^2
\end{equation}
It is crucial that this relation does not depend on the spin of the
massless fermion so that the whole infinite set of massless fields
with different spins can simultaneously interact with one and the same
massive supermultiplet $(\frac{1}{2},0)$. Recall that the same
relation we have found for this massive supermultiplet in
\cite{BSZ16}. 

Similarly, one can consider interaction for massless boson with two
massive bosons (see Appendix B) or two massive fermions. Let us
provide here a couple of explicit examples.

For the massless spin 2 and two massive bosons with arbitrary masses
and spins one can consider
\begin{eqnarray}
0 &=& D W_1^{\alpha(2k)} + e_{\beta(2)} W_1^{\alpha(2k)\beta(2)}
+ \alpha_{1,k} e^\alpha{}_\beta W_1^{\alpha(2k-1)\beta} +
\beta_{1,k} e^{\alpha(2)} W_1^{\alpha(2k-2)} \nonumber \\
 && + f_{k,0} \Omega_{\beta(2)} W_2^{\alpha(2k)\beta(2)}
+ f_{k,1} \Omega^\alpha{}_\beta W_2^{\alpha(2k-1)\beta}
+ f_{k,2} \Omega^{\alpha(2)} W_2^{\alpha(2k-2)} 
\end{eqnarray}
and similarly for the second boson. We again obtain the same relations
with $\Delta = 0,\pm 1$, where\\
$\Delta = 0$:
\begin{equation}
f_{k,0} = f_0, \qquad f_{k,1} = \alpha_k + \frac{\lambda}{2}, \qquad
f_{k,2} = \beta_k;
\end{equation}
$\Delta = \pm 1$:
\begin{eqnarray}
f_{k,0} &=& f_0, \qquad
f_{k,1} = \mp \frac{(k \mp s_1)}{2k(k+1)} [ M_1 \mp k\lambda]f_0,
\nonumber \\
f_{k,2} &=& \frac{(k \mp s_1-1)(k \mp s_1)}{2k^2(4k^2-1)}
[M_1 \mp k\lambda][M_1 \mp (k-1)\lambda]f_0.
\end{eqnarray}
We have found a complete solution for the massless spin 3  and indeed
found $\Delta = 0, \pm 1, \pm 2$.

\subsection{Three massive fields}

In this subsection we consider cubic vertices for three massive
fields. Again we begin with the purely bosonic case. First of all
note, that here we have a number of possible field redefinitions
$$
\Omega_1 \Rightarrow \Omega_1 + \kappa_1 \Omega_2 B_3 + \kappa_2 B_2
\Omega_3
$$
and similarly for two other one-forms. As in the previous case, using
these redefinitions one can completely remove all possible non-abelian
corrections to the gauge transformations. Further, any abelian vertex
of the form ${\cal B}{\cal B}\Omega$ by using the substitution
$$
\Omega^{\alpha(2s-2)} \Rightarrow - \frac{1}{\tilde{m}}
[ {\cal B}^{\alpha(2s-2)} - D B^{\alpha(2s-2)} 
- \frac{M}{2(s-1)} e^\alpha{}_\beta B^{\alpha(2s-3)\beta}]
$$
and differential identities for the curvatures can be shown to be
equivalent to the combination of the trivially gauge invariant ones
and abelian vertices of the form ${\cal B}{\cal B} B$, which can not
be made gauge invariant. This leaves us with the only possibility ---
trivially gauge invariant vertex:
\begin{equation}
{\cal L}_1 = g {\cal B}_1^{\alpha(\hat{s}_3)\beta(\hat{s}_2)}
{\cal B}_2^{\gamma(\hat{s}_1)}{}_{\alpha(\hat{s}_3)}
{\cal B}_{3,\beta(\hat{s}_2)\gamma(\hat{s}_1)}
\end{equation}
Similarly, for two massive fermions and one massive boson we can
construct:
\begin{equation}
{\cal L}_1 = g {\cal C}_1^{\alpha(\hat{s}_3+1)\beta(\hat{s}_2)}
{\cal C}_2^{\gamma(\hat{s}_1)}{}_{\alpha(\hat{s}_3+1)}
{\cal B}_{3,\beta(\hat{s}_2)\gamma(\hat{s}_1)}
\end{equation}
In both cases a strict triangular inequality must hold. Each term in
these vertices are separately gauge invariant so one can freely
consider both $+$ and $-$ components combining them into parity 
even/odd combinations similarly to the massless case.

An important remaining open question is the relation of such vertices
with the classification obtained recently in \cite{Met20} (see also
\cite{STT20}). 

\section{Partially massless fields}

To discuss the partially massless fields let us return back to the
initial Lagrangian (\ref{gm}) for massive boson. It is unitary not
only in $AdS_3$ and flat Minkowski spaces, but in $dS_3$ space as
well, provided
\begin{equation}
M^2 = m^2 - (s-1)^2 \Lambda \ge 0.
\end{equation}
At the boundary of the unitarity region lives the only partially
massless field which has one physical degrees of freedom. Inside the
region we find a number of partially massless fields described by a
set of one-forms $\Omega^{\alpha(2k)}$, $f^{\alpha(2k)}$, 
$s-1-t \le k \le s-1$, where $t$ is a depth of partially masslessness
(we define depth so that $t=0$ corresponds to the massless case).
These fields do not have any physical degrees of freedom and must be
considered separately.

\subsection{Maximal depth}

Here $M=0$ and zero-forms $\pi^{\alpha(2)}$ and
$\varphi$ decouple. The Lagrangian becomes:
\begin{eqnarray}
{\cal L}_0 &=& \sum_{k=1} (-1)^{k+1} [ k \Omega_{\alpha(2k-1)\beta}
e^\beta{}_\gamma \Omega^{\alpha(2k-1)\gamma} + \Omega_{\alpha(2k)}
D f^{\alpha(2k)} ] \nonumber \\
 && + E B_{\alpha(2)} B^{\alpha(2)} - B_{\alpha(2)} e^{\alpha(2)}
D A \nonumber \\
 && + \sum_{k=1} (-1)^{k+1} a_k [ - \frac{(k+2)}{k}
\Omega_{\alpha(2k)\beta(2)} e^{\beta(2)} f^{\alpha(2k)} + 
\Omega_{\alpha(2k)} e_{\beta(2)} f^{\alpha(2k)\beta(2)} ] \nonumber \\
 && + 2a_0 \Omega_{\alpha(2)} e^{\alpha(2)} A - a_0 
f_{\alpha\beta} E^\beta{}_\gamma B^{\alpha\gamma}, 
\end{eqnarray}
where
\begin{eqnarray}
a_k{}^2 &=& \frac{k(k+1)(s+k+1)(s-k-1)}{2(k+2)(2k+3)} \Lambda,
\nonumber \\
a_0{}^2 &=& \frac{(s+1)(s-1)}{3} \Lambda.
\end{eqnarray}
The gauge transformations are the same as before (but without $\pi$
and $\varphi$). The gauge invariant two-forms now:
\begin{eqnarray}
{\cal R}^{\alpha(2k)} &=& D \Omega^{\alpha(2k)} + \frac{(k+2)}{k}
a_k e_{\beta(2)} \Omega^{\alpha(2k)\beta(2)} + \frac{a_{k-1}}{k(2k-1)}
e^{\alpha(2)} \Omega^{\alpha(2k-2)}, \nonumber \\
{\cal T}^{\alpha(2k)} &=& D f^{\alpha(2k)} + a_k e_{\beta(2)}
f^{\alpha(2k)\beta(2)} + \frac{(k+1)a_{k-1}}{k(k-1)(2k-1)}
e^{\alpha(2)} f^{\alpha(2k-2)} + e^\alpha{}_\beta 
\Omega^{\alpha(2k-1)\beta}, \nonumber \\
{\cal R}^{\alpha(2)} &=& D \Omega^{\alpha(2)} + 3a_1
e_{\beta(2)} \Omega^{\alpha(2)\beta(2)},  \\
{\cal T}^{\alpha(2)} &=& D f^{\alpha(2)} + e^\alpha{}_\beta
\Omega^{\alpha\beta} + a_1 e_{\beta(2)} f^{\alpha(2)\beta(2)}
+ 2a_0 e^{\alpha(2)} A, \nonumber \\
{\cal R} &=& D A + \frac{a_0}{4} e_{\alpha(2)} f^{\alpha(2)} 
- E_{\alpha(2)} B^{\alpha(2)}. \nonumber
\end{eqnarray}
As for the one-forms, we obtain:
\begin{eqnarray}
{\cal B}^{\alpha(2)} &=& D B^{\alpha(2)} - 2a_0 \Omega^{\alpha(2)} 
+ 3a_1 e_{\beta(2)} B^{\alpha(2)\beta(2)}, \nonumber \\
{\cal B}^{\alpha(2k)} &=& D B^{\alpha(2k)} - 2a_0 \Omega^{\alpha(2k)}
 + \frac{a_{k-1}}{k(2k-1)} e^{\alpha(2)} B^{\alpha(2k-2)}
 + \frac{(k+2)}{k}a_k e_{\beta(2)} B^{\alpha(2k)\beta(2)}, \\
{\cal B}^{\alpha(2s-2)} &=& D B^{\alpha(2s-2)} - 2a_0
\Omega^{\alpha(2s-2)} + \frac{a_{s-2}}{(s-1)(2s-3)} e^{\alpha(2)}
B^{\alpha(2s-4)}. \nonumber
\end{eqnarray}
There also exists a set of self-consistent unfolded equations
\cite{Zin15}, namely, all curvatures except ${\cal B}^{\alpha(2s-2)}$
are zero, while
\begin{eqnarray}
0 &=& D B^{\alpha(2s-2)} - 2a_0
\Omega^{\alpha(2s-2)} + \frac{a_{s-2}}{(s-1)(2s-3)} e^{\alpha(2)}
B^{\alpha(2s-4)} + e_{\beta(2)} W^{\alpha(2s-2)\beta(2)}, \nonumber \\
0 &=& D W^{\alpha(2k)} + e_{\beta(2)} W^{\alpha(2k)\beta(2)}
+ \beta_k e^{\alpha(2)} W^{\alpha(2k-2)},
\end{eqnarray}
where
\begin{equation}
\beta_k = - \frac{(k^2-s^2)}{2(4k^2-1)} \Lambda.
\end{equation}

\noindent
{\bf Two partially massless and one massless fields.} First of all let
us note that in this case we have twice as many one-forms as
Stueckelberg zero-forms so our trick with maximal gauge fixing can not
be straightforwardly applied here. Leaving the general analysis to the
future work, we assume that in this case non-abelian corrections can
also be transformed into abelian ones by appropriate field
redefinitions and consider the following ansatz:
\begin{equation}
{\cal L}_1 = g {\cal B}_1^{\alpha(\hat{s}_3)\beta(\hat{s}_2)}
{\cal B}_2^{\gamma(\hat{s}_1)}{}_{\alpha(\hat{s}_3)}
f_{3,\beta(\hat{s}_2)\gamma(\hat{s}_1)},
\end{equation}
where $\hat{s}_i$ are the same  as before and the spins $s_i$ again
must satisfy a strict triangular inequality. Taking into account that
$$
D {\cal B}^{\alpha(2s-2)} = - 2a_0 {\cal R}^{\alpha(2s-2)}
 - \frac{a_{s-2}}{(s-1)(2s-3)} e^{\alpha(2)} {\cal B}^{\alpha(2s-4)}
\approx 0
$$
the only non-zero on-shell contribution appears to be:
\begin{equation}
\delta {\cal L}_1 \sim 
{\cal B}_1^{\alpha(\hat{s}_3)\beta(\hat{s}_2)}
{\cal B}_2^{\gamma(\hat{s}_1)}{}_{\alpha(\hat{s}_3)} 
( e_\beta{}^\delta \eta_{\beta(\hat{s}_2-1)\delta\gamma(\hat{s}_1)} +
e_\gamma{}^\delta \eta_{\beta(\hat{s}_2)\gamma(\hat{s}_1-1)\delta})
\end{equation}
From the unfolded equations it follows that ${\cal B}^{\alpha(2s-2)}
\approx e_{\beta(2)} W^{\alpha(2s-2)\beta(2)}$ so we obtain:
\begin{equation}
\delta {\cal L}_1 \sim (s_1-s_2) 
W_1^{\alpha(\hat{s}_3+2)\beta(\hat{s}_2)}
W_2^{\gamma(\hat{s}_1)}{}_{\alpha(\hat{s}_3+2)}
\eta_{\beta(\hat{s}_2)\gamma(\hat{s}_1)}.
\end{equation}
Thus gauge invariance requires that the spins of the two partially
massless fields must be equal $s_1 = s_2$. Indeed, this is just
the particular case of the general relation (\ref{gen_rel}) when
$M_1 = M_2 = 0$. Anyway, the result is rather strong, so as an
independent check we again consider the deformation of the unfolded
equations for the partially massless fields due to the massless
spin-$s$ field satisfying ($n = 2s-2$)
\begin{eqnarray}
0 &=& D \Omega^{\alpha(n)} - \frac{\Lambda}{4} e^\alpha{}_\beta
f^{\alpha(n-1)\beta} \nonumber \\
0 &=& D f^{\alpha(n)} + e^\alpha{}_\beta \Omega^{\alpha(n-1)\beta}
\end{eqnarray}
The most general ansatz for the parity even deformation has the form:
\begin{eqnarray}
0 &=& D W_1^{\alpha(2k)} + e_{\beta(2)} W_1^{\alpha(2k)\beta(2)}
+ \beta_{1,k} e^{\alpha(2)} W_1^{\alpha(2k-2)} \nonumber \\
 && + \sum_{l=0}^{n/2} f_{k,l} f^{\alpha(2l)}{}_{\beta(n-2l)}
W_2^{\alpha(2k-2l)\beta(n-2l)} \nonumber \\
 && + \sum_{l=0}^{n/2-1} g_{k,l}
\Omega^{\alpha(2l+1)}{}_{\beta(n-2l-1)}
W_2^{\alpha(2k-2l-1)\beta(n-2l-1)} 
\end{eqnarray}
Self-consistency of such deformations provide a large number of
equations on the parameters $f_{k,l}$ and $g_{k,l}$ which can be
solved iteratively on $l$ (see Appendix C). For $l=0$ they give:
$$
f_{k+1,0} = f_{k,0}, \qquad
g_{k+1,0} = g_{k,0}, \qquad
g_{k+1,0} = \frac{n}{2}f_{k,0}
$$
so we set 
$$
f_{k,0} = f_0 = 1, \qquad g_{k,0} = \frac{n}{2}
$$
Equations with $l=1$ for the field $f$ look like:
\begin{eqnarray*}
0 &=& 2f_{k+1,1} - (2k-1)\beta_{1,k} + (2k-1+n)\beta_{2,k+n/2} +
\frac{n^2}{8}\Lambda  \\
0 &=& f_{k+1,1} + k(2k-1)\beta_{1,k} 
- \frac{(2k+n)(2k+n-1)}{2}\beta_{2,k+n/2} - \frac{kn}{4}\Lambda \\
0 &=& f_{k+1,1} - f_{k,1} + \beta_{1,k} - \beta_{2,k+n/2} 
\end{eqnarray*}
For $f_{k,1}$ from the first equation the second one satisfies
identically, while the third one requires
$$
s_1 = s_2
$$
Similarly, equations for $\Omega$ with $l=1$ look like
\begin{eqnarray*}
0 &=& g_{k+1,1} - n(k-1)\beta_{1,k} + n(2k-3+n)\beta_{2,k-1+n/2}/2 -
\frac{n}{2}f_{k,1} \\
0 &=& g_{k+1,1} - g_{k,1} + \frac{3n}{2}(\beta_{1,k} - 
\beta_{2,k-1+n/2}) \\
0 &=& g_{k+1,1} + \frac{n(k-1)(2k-1)}{2}\beta_{1,k} 
- \frac{(2k+n-2)(2k+n-3)}{4} \beta_{2,k-1+n/2} + (2k-1)f_{k,1} 
\end{eqnarray*}
and again solution exists only when $s_1 = s_2$. Let us provide here
complete solutions for a couple of simple cases.

\noindent
{\bf Massless spin 2} In this case equations are:
\begin{eqnarray}
0 &=& D W_1^{\alpha(2k)} + e_{\beta(2)} W_1^{\alpha(2k)\beta(2)}
+ \beta_k e^{\alpha(2)} W_1^{\alpha(2k-2)} \nonumber \\
 && + f_{k,0} f_{\beta(2)} W_2^{\alpha(2k)\beta(2)} 
+ g_{k,0} \Omega^\alpha{}_\beta W_2^{\alpha(2k-1)\beta}
+ f_{k,1} \Phi^{\alpha(2)} W_2^{\alpha(2k-2)}
\end{eqnarray}
and the solution:
$$
f_{k,0} = g_{k,0} = 1, \qquad f_{k,1} = \beta_k
$$
{\bf Massless spin 3} Now we have
\begin{eqnarray}
0 &=& D W_1^{\alpha(2k)} + e_{\beta(2)} W_1^{\alpha(2k)\beta(2)}
+ \beta_k e^{\alpha(2)} W_1^{\alpha(2k-2)} \nonumber \\
 && + f_{k,0} f_{\beta(4)} W_2^{\alpha(2k)\beta(4)}
+ f_{k,1} f^{\alpha(2)}{}_{\beta(2)} W_2^{\alpha(2k-2)\beta(2)}
+ f_{k,2} f^{\alpha(4)} W_3^{\alpha(2k-4)} \nonumber \\
 && + g_{k,0} \Omega^\alpha{}_{\beta(3)} W_2^{\alpha(2k-1)\beta(3)}
+ g_{k,1} \Omega^{\alpha(3)}{}_\beta W_2^{\alpha(2k-3)\beta}
\end{eqnarray}
and obtain:
$$
f_{k,0} = f_0 = 1, \qquad
f_{k,1} = 2[(k+2)\beta_{k+1} - (k-1)\beta_k] , \qquad
f_{k,2} = 6\beta_k\beta_{k-1}
$$
$$
g_{k,0} = g_0 = 2f_0, \qquad
g_{k,1} = 6\beta_k
$$

\subsection{Non-maximal depth}

Recently, a very simple and elegant representation for partially
massless fields of non-maximal depth in $AdS_3$ were suggested
\cite{GMS20}. The aim of this subsection is to show that this
representation can be reproduced starting with our gauge invariant
formalism. First of all we have to move to $AdS$. These partially
massless fields are non-unitary there, so if we formally set $\Lambda
= - \lambda^2$ we obtain non-hermitian Lagrangian. To avoid this, we
use the same trick as we have already used in our construction for the
partially massless supermultiplets in $AdS_4$ \cite{BKhSZ19a}. Namely,
we switch off alternating signs for the kinetic terms. Then we obtain
the Lagrangian ($n=s-t-1$):
\begin{eqnarray}
{\cal L}_0 &=& \sum_{k=n}^{s-1} [ k \Omega_{\alpha(2k-1)\beta}
e^\beta{}_\gamma \Omega^{\alpha(2k-1)\gamma} + \Omega_{\alpha(2k)}
D f^{\alpha(2k)} + b_k f_{\alpha(2k-1)\beta} e^\beta{}_\gamma
f^{\alpha(2k-1)\gamma} \nonumber \\
 && \qquad - \frac{(k+2)}{k}a_k \Omega_{\alpha(2k)\beta(2)}
e^{\beta(2)} f^{\alpha(2k)} + a_k \Omega_{\alpha(2k)}
e_{\beta(2)} f^{\alpha(2k)\beta(2)} ], 
\end{eqnarray}
where all coefficients are real and such that
$$
a_k{}^2 = \frac{k(s^2-(k+1)^2)((k+1)^2-n^2)}{2(k+1)(k+2)(2k+3)}
\lambda^2, \qquad
b_k = \frac{s^2n^2}{4k(k+1)^2} \lambda^2.
$$
This Lagrangian is invariant under the following gauge
transformations:
\begin{eqnarray*}
\delta \Omega^{\alpha(2k)} &=& D \eta^{\alpha(2k)} - \frac{(k+2)}{k}
a_k e_{\beta(2)} \eta^{\alpha(2k)\beta(2)} + 
\frac{a_{k-1}}{k(2k-1)} e^{\alpha(2)} \eta^{\alpha(2k-2)}
+ \frac{b_k}{k} e^\alpha{}_\beta \xi^{\alpha(2k-1)\beta}, \\
\delta f^{\alpha(2k)} &=& D \xi^{\alpha(2k)} - a_k e_{\beta(2)}
\xi^{\alpha(2k)\beta(2)} + \frac{(k+1)a_{k-1}}{k(k-1)(2k-1)}
e^{\alpha(2)} \xi^{\alpha(2k-2)} + e^\alpha{}_\beta
\eta^{\alpha(2k-1)\beta}, 
\end{eqnarray*}
while the gauge invariant two-forms look like:
\begin{eqnarray*}
{\cal R}^{\alpha(2k)} &=& D \Omega^{\alpha(2k)} - \frac{(k+2)}{k}
a_k e_{\beta(2)} \Omega^{\alpha(2k)\beta(2)} + 
\frac{a_{k-1}}{k(2k-1)} e^{\alpha(2)} \Omega^{\alpha(2k-2)} +
\frac{b_k}{k} e^\alpha{}_\beta f^{\alpha(2k-1)\beta}, \\
{\cal T}^{\alpha(2k)} &=& D f^{\alpha(2k)} + e^\alpha{}_\beta
\Omega^{\alpha(2k-1)\beta} - a_k e_{\beta(2)}
f^{\alpha(2k)\beta(2)} + \frac{(k+1)a_{k-1}}{k(k-1)(2k-1)}
e^{\alpha(2)} f^{\alpha(2k-2)}.
\end{eqnarray*}
Now to separate the variables we introduce:
\begin{equation}
\Omega_\pm^{\alpha(2k)} = \Omega^{\alpha(2k)} \pm
\frac{sn\lambda}{2k(k+1)} f^{\alpha(2k)}
\end{equation}
Working with the $+$ components and omitting the $+$ sign we obtain:
\begin{eqnarray}
{\cal R}^{\alpha(2k)} &=& D \Omega^{\alpha(2k)} - \frac{(k+2)}{k}a_k
e_{\beta(2)} \Omega^{\alpha(2k)\beta(2)} \nonumber \\
 && + \frac{a_{k-1}}{k(2k-1)} e^{\alpha(2)} \Omega^{\alpha(2k-2)} +
\frac{sn\lambda}{2k(k+1)} e^\alpha{}_\beta \Omega^{\alpha(2k-1)\beta} 
\end{eqnarray}

As an illustration let us consider the simplest case $t=1$. In this
case we have only two fields $\Omega^{\alpha(2s-2)}$ and
$\Omega^{\alpha(2s-4)}$ with curvatures
\begin{eqnarray}
{\cal R}^{\alpha(2s-2)} &=& D \Omega^{\alpha(2s-2)} 
+ \frac{2a_{s-2}}{(s-1)(2s-3)} \frac{\lambda}{2} e^{\alpha(2)}
\Omega^{\alpha(2s-4)} + \frac{(s-2)}{(s-1)} \frac{\lambda}{2}
e^\alpha{}_\beta \Omega^{\alpha(2s-3)\beta}, \nonumber \\
{\cal R}^{\alpha(2s-4)} &=& D \Omega^{\alpha(2s-4)} - 
\frac{2sa_{s-2}}{(s-2)} \frac{\lambda}{2} e_{\beta(2)}
\Omega^{\alpha(2s-4)\beta(2)} + \frac{s}{(s-1)} \frac{\lambda}{2}
e^\alpha{}_\beta \Omega^{\alpha(2s-5)\beta},
\end{eqnarray}
where we explicitly show multipliers $\frac{\lambda}{2}$ so that
\begin{equation}
a_{s-2}{}^2 = \frac{(s-2)(2s-3)}{2s(s-1)}.
\end{equation}
Now we show that both these curvatures can be combined into
\begin{equation}
\hat{\cal R}^{\alpha(2s-3),\beta} = D \hat\Omega^{\alpha(2s-3)\beta}
+ \frac{\lambda}{2} e^\alpha{}_\gamma 
\hat\Omega^{\alpha(2s-4)\gamma,\beta} - \frac{\lambda}{2}
e^\beta{}_\gamma \hat\Omega^{\alpha(2s-3),\gamma}.
\end{equation}
For this we introduce an ansatz:
\begin{equation}
\hat\Omega^{\alpha(2s-3),\beta} = \Omega^{\alpha(2s-3)\beta}
+ \kappa_1 \Omega^{\alpha(2s-4)} \varepsilon^{\alpha\beta}
\end{equation}
Let us begin with the $\Omega^{\alpha(2s-2)}$ contribution (here and
further on all calculations are up to $\frac{\lambda}{2}$):
\begin{equation}
\Delta_0 = e^\alpha{}_\gamma \Omega^{\alpha(2s-4)\beta\gamma} -
e^\beta{}_\gamma \Omega^{\alpha(2s-3)\gamma}
\end{equation}
Using an identity:
$$
e^\alpha{}_\gamma \Omega^{\alpha(2s-4)\beta\gamma} - (2s-3)
e^\beta{}_\gamma \Omega^{\alpha(2s-3)\gamma} =
\varepsilon^{\alpha\beta} e_{\gamma(2)} 
\Omega^{\alpha(2s-4)\gamma(2)},
$$
we can rewrite this contribution as
\begin{equation}
\Delta_0 = \frac{(s-2)}{(s-1)} [ e^\alpha{}_\gamma
\Omega^{\alpha(2s-4)\beta\gamma} + e^\beta{}_\gamma
\Omega^{\alpha(2s-3)\gamma}] + \frac{1}{(s-1)}
\varepsilon^{\alpha\beta} e_{\gamma(2)} 
\Omega^{\alpha(2s-4)\gamma(2)}.
\end{equation}
The coefficient before the first term is exactly what we need, while
for the second contribution be correct we must have
\begin{equation}
\frac{1}{(s-1)} = - \kappa_1 \frac{2s}{(s-2)} a_{s-2} \quad
\Longrightarrow \quad \kappa_1 = - \frac{a_{s-2}}{(2s-3)}.
\end{equation}
Now we consider $\Omega^{\alpha(2s-4)}$ contribution:
\begin{equation}
\Delta_1 = e^\alpha{}_\gamma \Omega^{\alpha(2s-5)\gamma}
\varepsilon^{\alpha\beta} - 2 e^{\alpha\beta} \Omega^{\alpha(2s-4)}
\end{equation}
Using one more identity:
$$
e^\alpha{}_\gamma \Omega^{\alpha(2s-5)\gamma}
\varepsilon^{\alpha\beta} = 2 e^{\alpha(2)} \Omega^{\alpha(2s-5)\beta}
- 2(s-2) e^{\alpha\beta} \Omega^{\alpha(2s-4)},
$$
we obtain
\begin{equation}
\Delta_1 = \frac{s}{(s-1)} e^\alpha{}_\gamma 
\Omega^{\alpha(2s-5)\gamma} \varepsilon^{\alpha\beta} - 
\frac{2}{(s-1)} ( e^{\alpha(2)} \Omega^{\alpha(2s-5)\beta} +
e^{\alpha\beta} \Omega^{\alpha(2s-4)}).
\end{equation}
Taking into account that
$$
- \kappa_1 \frac{2}{(s-1)} = \frac{2a_{s-2}}{(s-1)(2s-3)}
$$
we find that both coefficients are correct.

Now we turn to the general case and show that the whole set of the
gauge invariant curvatures can be packed into
\begin{equation}
\hat{\cal R}^{\alpha(2s-2-t),\beta(t)} = D
\hat\Omega^{\alpha(2s-2-t),\beta(t)} + \frac{\lambda}{2}
e^\alpha{}_\gamma \hat\Omega^{\alpha(2s-3-t)\gamma,\beta(t)} -
\frac{\lambda}{2} e^\beta{}_\gamma
\hat\Omega^{\alpha(2s-2-t),\beta(t-1)\gamma}.
\end{equation}
For this purpose we introduce the following ansatz:
\begin{equation}
\hat\Omega^{\alpha(2s-2-t),\beta(t)} = \Omega^{\alpha(2s-2-t)\beta(t)}
+ \sum_{l=1}^t \kappa_l \Omega^{\alpha(2s-2-t-l)\beta(t-l)}
(\varepsilon^{\alpha\beta})^l,
\end{equation}
where $(\varepsilon^{\alpha\beta})^l$ denotes a product of $l$ copies
of $\varepsilon^{\alpha\beta}$. It is enough to consider a
contribution for one particular field  $\Omega^{\alpha(2s-2-2l)}$. For
convenience, let us explicitly show the terms we have to reproduce:
\begin{eqnarray}
{\cal R}^{\alpha(2s-2l)} &=& \dots + \frac{2a_{s-l-1}}{(s-l)(2s-2l-1)}
e^{\alpha(2)} \Omega^{\alpha(2s-2-2l)} + \dots \label{c1} \\
{\cal R}^{\alpha(2s-2-2l)} &=& \dots + \frac{s(s-t-1)}{(s-l)(s-l-1)}
e^\alpha{}_\beta \Omega^{\alpha(2s-3-2l)\beta} \label{c2} \\
{\cal R}^{\alpha(2s-4-2l)} &=& \dots - \frac{2(s-l)}{(s-l-2)}
a_{s-l-2} e_{\beta(2)} \Omega^{\alpha(2s-4-2l)\beta(2)} + \dots
\label{c3}
\end{eqnarray}
This particular contribution appears to be:
\begin{eqnarray}
\Delta_l &=& [ e^\alpha{}_\gamma 
\Omega^{\alpha(2s-3-t-l)\gamma\beta(t-l)} - e^\beta{}_\gamma
\Omega^{\alpha(2s-2-t-l)\beta(t-l-1)\gamma}]
(\varepsilon^{\alpha\beta})^l \nonumber \\
 && - 2l e^{\alpha\beta} \Omega^{\alpha(2s--2-t-l)\beta(t-l)}
(\varepsilon^{\alpha\beta})^{l-1}
\end{eqnarray}
Further on we omit common multiplier 
$(\varepsilon^{\alpha\beta})^{l-1}$. Using a pair of identities
\begin{eqnarray*}
e^\alpha{}_\gamma \Omega^{\alpha(2s-3-t-l)\gamma\beta(t-l)} 
\varepsilon^{\alpha\beta} &=& 2(t-l+1) e^{\alpha(2)}
\Omega^{\alpha(2s-3-t-l)\beta(t-l+1)} \\
 && - (2s-2-t-l) e^{\alpha\beta} 
\Omega^{\alpha(2s-2-t-l)\beta(t-l)}, \\
e^\beta{}_\gamma \Omega^{\alpha(2s-2-t-l)\beta(t-l-1)\gamma}
\varepsilon^{\alpha\beta} &=& - 2(2s-1-t-l) e^{\beta(2)}
\Omega^{\alpha(2s-1-t-l)\beta(t-l-1)} \\
 && + (t-l) e^{\alpha\beta} \Omega^{\alpha(2s-2-t-l)\beta(t-l)},
\end{eqnarray*}
one can straightforwardly show that
\begin{eqnarray}
\Delta_{l,1} &=& [\rho_1 e^\alpha{}_\gamma 
\Omega^{\alpha(2s-3-t-l)\beta(t-l)\gamma} + \rho_2 e^\beta{}_\gamma
\Omega^{\alpha(2s-2-t-l)\beta(t-l-1)\gamma} ]
\varepsilon^{\alpha\beta} \nonumber \\
 && - 2l e^{\alpha\beta} \Omega^{\alpha(2s-2-t-l)\beta(t-l)}.
\end{eqnarray}
where
$$
\rho_1 = - \frac{l(2s-t-l-1)}{(s-l)(2s-2l-1)}, \qquad
\rho_2 = \frac{l(t-l+1)}{(s-l)(2s-2l-1)},
$$
can be rewritten as:
\begin{eqnarray}
\Delta_{l,1} &=& - \frac{l(2s-t-l-1)}{(s-l)(2s-2l-1)}
[ e^{\alpha(2)} \Omega^{\alpha(2s-3-t-l)\beta(t-l+1)} \nonumber \\
 && \qquad\qquad + e^{\alpha\beta} \Omega^{\alpha(2s-2-t-l)\beta(t-l)}
+ e^{\beta(2)} \Omega^{\alpha(2s-1-t-l)\beta(t-l-1)} ].
\end{eqnarray}
To correctly reproduce the term in (\ref{c1}) we must have 
$$
- \kappa_l \frac{l(2s-t-l-1)}{(s-l)(2s-2l-1)} = \kappa_{l-1}
\frac{2a_{s-l-1}}{(s-l)(2s-2l-1)}.
$$
This gives us a recurrent relation on $\kappa_l$:
\begin{equation}
\kappa_l = - \kappa_{l-1} \frac{a_{s-l-1}}{l(t-l+1)(2s-t-l-1)}.
\end{equation}
Taking into account that $\kappa_0 = 1$ we found a solution
\begin{equation}
\kappa_l = (-1)^l \frac{(t-l)!(2s-2-t-l)!}{l!t!(2s-2-t)!}
\prod_{m=1}^l a_{s-1-m}.
\end{equation}
This leaves us with (omitting the last $\varepsilon^{\alpha\beta}$)
$$
\Delta_{l,2} = (1-\rho_1) e^\alpha{}_\gamma
\Omega^{\alpha(2s-3-t-l)\gamma\beta(t-l)} - (1+\rho_2)e^\beta{}_\gamma
\Omega^{\alpha(2s-2-t-l)\beta(t-l-1)\gamma}.
$$
Using the last identity
\begin{eqnarray*}
(t-l)e^\alpha{}_\gamma \Omega^{\alpha(2s-3-t-l)\gamma\beta(t-l)}
&=& (2s-2-t-l) e^\beta{}_\gamma 
\Omega^{\alpha(2s-2-t-l)\beta(t-l-1)\gamma} \\
 && + \varepsilon^{\alpha\beta} e_{\gamma(2)}
\Omega^{\alpha(2s-3-t-l)\beta(t-l-1)\gamma(2)},
\end{eqnarray*}
we obtain:
\begin{eqnarray*}
\Delta_{l,2} &=& \frac{s(s-t-1)}{(s-l)(s-l-1)}
[ e^\alpha{}_\gamma \Omega^{\alpha(2s-3-t-l)\gamma\beta(t-l)} +
e^\beta{}_\gamma \Omega^{\alpha(2s-2-t-l)\beta(t-l-1)\gamma} ] \\
 && + \frac{(2s-l-1)}{(s-l-1)(2s-2l-1)} \varepsilon^{\alpha\beta}
e_{\gamma(2)} \Omega^{\alpha(2s-3-t-l)\beta(t-l-1)\gamma(2)}.
\end{eqnarray*}
Here the first coefficients is exactly what we need for (\ref{c2}),
while to reproduce (\ref{c3}) we must have
$$
\kappa_l \frac{(2s-l-1)}{(s-l-1)(2s-2l-1)} =
- \kappa_{l+1} \frac{2(s-l)}{(s-l-2)} a_{s-l-2}.
$$
Using explicit expressions for $\kappa_l$ and $a_{s-2}$ it is
straightforward to check that this equation is fulfilled.

\section*{Acknowledgments} 

Author is grateful to I. L. Buchbinder and  T. V. Snegirev for
collaboration.

\appendix

\section{Massless fermion and massive boson and fermion}

Let us consider massive boson $(M_1,s_1)$ and massive fermion 
$(M_2,s_2)$ interacting with massless spin-$s$ fermion satisfying
$$
0 = D \Phi^{\alpha(n)} + \frac{\lambda}{2} e^\alpha{}_\beta
\Phi^{\alpha(n-1)\beta}, \qquad n = 2s-2
$$
Deformation for the bosonic equations has the form:
\begin{eqnarray}
0 &=& D W^{\alpha(2k)} + e_{\beta(2)} W^{\alpha(2k)\beta(2)}
+ \alpha_{1,k} e^\alpha{}_\beta W^{\alpha(2k-1)\beta} 
+ \beta_{1,k} e^{\alpha(2)} W^{\alpha(2k-2)} \nonumber \\
 && + \sum_{l=0}^n f_{k,l} \Phi^{\alpha(l)}{}_{\beta(n-l)}
V^{\alpha(2k-l)\beta(n-l)} 
\end{eqnarray}
The consistency requirement leads to the number of equations which can
(in-principle) be solved iteratively by $l$:\\
$l=0$
\begin{eqnarray*}
0 &=& f_{k+1,0} - f_{k,0} \quad \Rightarrow \quad f_{k,0} = f_0 \\
0 &=& 2f_{k+1,1} + (2k+n) \alpha_{2,k+(n-1)/2} - 2k\alpha_{1,k}
- \frac{n\lambda}{2} \\
0 &=& f_{k+1,1} - f_{k,1} + \alpha_{1,k} - \alpha_{2,k+(n-1)/2}
\end{eqnarray*}
$1 \le l \le n-1$:
\begin{eqnarray}
0 &=& 2f_{k+1,l+1} + [(2k-2l+n)\alpha_{2,k-l+(n-1)/2} -
(2k-2l)\alpha_{1,k} - n\frac{\lambda}{2}]f_{k,l} \nonumber \\
 && - l(2k-l)\beta_{1,k}f_{k-1,l-1} + l(2k-2l+n+1) 
\beta_{2,k-l+(n+1)/2}f_{k,l-1} \nonumber \\
0 &=& f_{k+1,l+1} - f_{k,l+1} 
+ (l+1) (\alpha_{1,k} - \alpha_{2,k-l+(n-1)/2})f_{k,l} \nonumber \\
 && + \frac{l(l+1)}{2} [\beta_{1,k}f_{k-1,l-1} -
\beta_{2,k-l+(n+1)/2}f_{k,l-1}] \\
0 &=& 2f_{k+1,l+1} - (2k-l+1)(2\alpha_{1,k}-\lambda)f_{k,l}
+ (2k-l)(2k-l+1)\beta_{1,k}f_{k-1,l-1} \nonumber \\
 && - (2k-2l+n+2)(2k-2l+n+1) \beta_{2,k-l+(n+1)/2}f_{k,l-1} \nonumber 
\end{eqnarray}
$l=n$:
\begin{eqnarray*}
0 &=& (\alpha_{1,k} - \alpha_{2,k-(n+1/2})f_{k,n} 
+ \frac{n}{2}[\beta_{1,k}f_{k-1,n-1} - \beta_{2,k-(n-1)/2}f_{k,n-1}] 
\\
0 &=& (2\alpha_{1,k}-\lambda)f_{k,n} + (2k-n)\beta_{1,k}f_{k-1,n-1}
- (2k-n+2)\beta_{2,k-(n-1)/2}f_{k,n-1} \\
0 &=& \beta_{1,k}f_{k-1,n} - \beta_{2,k-(n-1)/2}f_{k,n}
\end{eqnarray*}
Similarly, deformation for the fermionic equations can be written as:
\begin{eqnarray}
0 &=& D V^{\alpha(2k+1)} + e_{\beta(2)} V^{\alpha(2k+1)\beta(2)}
+ \alpha_{2,k} e^\alpha{}_\beta V^{\alpha(2k)\beta}
+ \beta_{2,k} e^{\alpha(2)} V^{\alpha(2k-1)} \nonumber \\
 && + \sum_{l-0}^n \Phi^{\alpha(l)}{}_{\beta(n-l)}
W^{\alpha(2k+1-l)\beta(n-l)}
\end{eqnarray}
and their consistency leads to:\\
$l=0$
\begin{eqnarray*}
0 &=& g_{k+1,0} - g_{k,0}, \quad \Rightarrow \quad g_{k,0} = g_0 \\
0 &=& g_{k+1,1} - g_{k,1} + \alpha_{2,k} - \alpha_{1,k+(n+1)/2} \\
0 &=& 2g_{k+1,1} + (2k+1+n)\alpha_{1,k+(n+1)/2} - (2k+1)\alpha_{2,k} 
- \frac{n\lambda}{2} 
\end{eqnarray*}
$1 \le l \le n-1$
\begin{eqnarray}
0 &=& 2g_{k+1,l+1} + [(2k+1-2l+n)\alpha_{1,k-l+(n+1)/2} 
- (2k+1-2l) \alpha_{2,k} - n\frac{\lambda}{2}]g_{k,l} \nonumber \\
 && + l(2k+2-2l+n)\beta_{1,k-l+(n+3)/2}g_{k,l-1}
- l(2k+1-l) \beta_{2,k}g_{k-1,l-1} \nonumber \\
0 &=& g_{k+1,l+1} - g_{k,l+1}
 + (l+1) (\alpha_{2,k} - \alpha_{1,k-l+(n+1)/2})g_{k,l} \nonumber \\
 && + \frac{l(l+1)}{2} [beta_{2,k}g_{k-1,l-1} 
- \beta_{1,k-l+(n+3)/2}g_{k,l-1}] \\
0 &=& 2g_{k+1,l+1} - (2k+2-l) (2\alpha_{2,k}-\lambda)g_{k,l}
+ (2k+2-l) [ (2k+1-l) \beta_{2,k}g_{k-1,l-1} \nonumber \\
 && - (2k-2l+n+3)(2k-2l+n+2) \beta_{1,k-l+(n+3)/2}g_{k,l-1}] \nonumber
\end{eqnarray}
$l=n$
\begin{eqnarray*}
0 &=& (\alpha_{2,k}-\alpha_{1,k-(n-1)/2})g_{k,n} 
+ \frac{n}{2}(\beta_{2,k}g_{k-1,n-1} - \beta_{1,k-(n-3)/2}g_{k,n-1}) 
\\
0 &=& (2\alpha_{2,k}-\lambda)g_{k,n} - (2k+1-n)\beta_{2,k}g_{k-1,n-1}
+ (2k+3-n)\beta_{1,k-(n-3)/2}g_{k,n-1} \\
0 &=& \beta_{2,k}g_{k-1,n} - \beta_{1,k-(n-1)/2}g_{k,n} 
\end{eqnarray*}

\section{One massless and two massive bosons}

Deformation for the first boson:
\begin{eqnarray}
0 &=& D W_1^{\alpha(2k)} + e_{\beta(2)} W_1^{\alpha(2k)\beta(2)}
+ \alpha_{1,k} e^\alpha{}_\beta W_1^{\alpha(2k-1)\beta} 
+ \beta_{1,k} e^{\alpha(2)} W_1^{\alpha(2k-2)} \nonumber \\
 && + \sum_{l=0}^n f_{k,l} \Omega^{\alpha(l)}{}_{\beta(n-l)}
W_2^{\alpha(2k-l)\beta(n-l)} 
\end{eqnarray}
Consistency requirement leads to:
\begin{eqnarray*}
0 &=& f_{k+1,0} - f_{k,0} \quad \Rightarrow \quad f_{k,0} = f_0 = 1 \\
0 &=& 2f_{k+1,1} + (2k+n)\alpha_{2,k+n/2} - 2k\alpha_{1,k} -
n\frac{\lambda}{2} \\ 
0 &=& f_{k+1,1} - f_{k,1} + \alpha_{1,k} - \alpha_{2,k+n/2}
\end{eqnarray*}
$1 \le l \le n-1$
\begin{eqnarray}
0 &=& 2f_{k+1,l+1} - (2k-l+1)(2\alpha_{1,k}-\lambda)f_{k,l}
+ (2k-l)(2k-l+1)\beta_{1,k}f_{k-1,l-1} \nonumber \\
 && - (2k-2l+n+2)(2k-2l+n+1) \beta_{2,k-l+1+n/2}f_{k,l-1} \nonumber \\
0 &=& f_{k+1,l+1} - f_{k,l+1} 
+ (l+1) [ \alpha_{1,k}f_{k,l} - \alpha_{2,k-l+n/2}f_{k,l} ] \nonumber
\\
 && + \frac{l(l+1)}{2} [\beta_{1,k} f_{k-1,l-1} -
\beta_{2,k-l+1+n/2}f_{k,l-1} ] \\
0 &=& 2f_{k+1,l+1} 
+ [(2k-2l+n)\alpha_{2,k-l+n/2} - (2k-2l)\alpha_{1,k} 
- n\frac{\lambda}{2}]f_{k,l} \nonumber \\
 && - l(2k-l)\beta_{1,k}f_{k-1,l-1} + l(2k-2l+n+1)
\beta_{2,k-l+1+n/2}f_{k,l-1} \nonumber
\end{eqnarray}
$l=n$
\begin{eqnarray*}
0 &=& (2\alpha_{1,k} - \lambda)f_{k,n} - (2k-n)\beta_{1,k}f_{k-1,n-1}
+ (2k-n+2)\beta_{2,k+1-n/2}f_{k,n-1} \\
0 &=& 2(\alpha_{1,k} - \alpha_{2,k-l+n/2})f_{k,n} 
+ n[\beta_{1,k}f_{k-1,n-1} - \beta_{2,k+1-n/2}f_{k,n-1}] \\
0 &=& \beta_{1,k}f_{k-1,n} - \beta_{2,k-n/2}f_{k,n} 
\end{eqnarray*}

\section{One massless and two partially massless bosons}

Deformation for the first partially massless one:
\begin{eqnarray}
0 &=& D W_1^{\alpha(2k)} + e_{\beta(2)} W_1^{\alpha(2k)\beta(2)}
+ \beta_{1,k} e^{\alpha(2)} W_1^{\alpha(2k-2)} \nonumber \\
 && + \sum_{l=0}^{n/2} f_{k,l} f^{\alpha(2l)}{}_{\beta(n-2l)}
 W_2^{\alpha(2k-2l)\beta(n-2l)} \nonumber \\
 && + \sum_{l=1}^{n/2-1} g_{k,l} 
\Omega^{\alpha(2l+1)}{}_{\beta(n-2l-1)}
 W_2^{\alpha(2k-2l-1)\beta(n-2l-1)} 
\end{eqnarray}
Terms with the $f^{\alpha(n)}$ field lead to:
$$
f_{k+1,0} = f_{k,0} \quad \Rightarrow \quad f_{k,0} = 0
$$
$1 \le l \le n/2$
\begin{eqnarray*}
0 &=& 2f_{k,l}  
- (2l-1) [(2k-2l-1)\beta_{1,k-1}f_{k-2,l-1} \\
 && \qquad - (2k-4l+1+n)\beta_{2,k-2l+1+n/2}f_{k-2,l-1}]
+ \frac{n\Lambda}{4}g_{k-1,l-1} \\
0 &=& f_{k+1l} - f_{k,l} 
+ l(2l-1) [\beta_{1,k}f_{k-1,l-1} - \beta_{2,k-2l+2+n/2}f_{k,l-1}] \\
0 &=& f_{k,l}  -
\frac{(2k-4l+n+2)(2k-4l+n+1))}{2}\beta_{2,k-2l+1+n/2}f_{k-1,l-1} \\
 && + (k-l) (2k-2l-1) \beta_{1,k-1}f_{k-2,l-1} -
(k-l)\frac{\Lambda}{2}g_{k-1,l-1}
\end{eqnarray*}
$$
0 = \beta_{1,k}f_{k-1,n/2} - \beta_{2,k-n/2}f_{k,n/2}
$$
Terms with the $\Omega^{\alpha(n)}$ field lead to:
$$
g_{k+1,0} - g_{k,0} = 0, \qquad g_{k+1,0} - \frac{n}{2}f_{k,0} = 0
$$
$1 \le l \le n/2-1$
\begin{eqnarray*}
0 &=& g_{k,l}  
- l [2(k-l-1)\beta_{1,k-1}g_{k-2,l-1} - (2k-4l-1+n)
\beta_{2,k-2l+n/2}g_{k-1,l-1}] - \frac{n}{2}f_{k-1,l} \\
0 &=& g_{k+1,l} - g_{k,l} 
+ l(2l+1)[\beta_{1,k}g_{k-1,l-1} - \beta_{2,k-2l+1+n/2}g_{k,l-1}] \\
0 &=& g_{k,l} -
\frac{(2k-4l+n)(2k-4l+n-1)}{2}\beta_{2,k-2l+n/2}g_{k-1,l-1} \\
 && + (2k-2l-1) (k-l-1)\beta_{1,k-1}g_{k-2,l-1} + (2k-2l-1) f_{k-1,l}
\end{eqnarray*}
$$
\beta_{1,k}g_{k-1,n/2-1} - \beta_{2,k-n/2+1}g_{k,n/2-1} = 0, \qquad
\beta_{2,k-n/2+1}g_{k,n/2-1} - f_{k,n/2} = 0
$$

\end{document}